\documentclass[useAMS,usenatbib,usegraphicx,letterpaper]{mn2e}
\usepackage[totalwidth=480pt, totalheight=680pt]{geometry}
\usepackage{ gensymb }

\newcommand{\p}{\partial}
\newcommand{\dd}{{\rm d}}

\newcommand{\sh}{{\rm sh}}
\newcommand{\M}{{\cal M}}

\usepackage{epsfig}
\usepackage{amsmath}
\usepackage{natbib}
\usepackage{verbatim}

\title[On the linear growth mechanism driving the stationary accretion shock instability]
{On the linear growth mechanism driving the stationary accretion shock instability}

\author[J\'er\^ome Guilet and Thierry Foglizzo]
{J\'er\^ome Guilet$^{1,2}$ \& Thierry Foglizzo$^{2}$ \\
$^{1}$ Department of Applied Mathematics and Theoretical Physics, University of Cambridge \\
Wilberforce Road, Cambridge CB3 0WA, UK. \\
$^{2}$ Laboratoire AIM, CEA/DSM-CNRS-Universit\'e Paris Diderot, IRFU/Service d'Astrophysique, \\
CEA-Saclay F-91191 Gif-sur-Yvette, France. 
}

\begin{document}

\maketitle

\label{firstpage}

\begin{abstract}
During stellar core collapse, which eventually leads to a supernovae explosion, the stalled shock is unstable due to  the standing accretion shock instability (SASI). This instability induces large-scale non spherical oscillations of the shock, which have crucial consequences on the dynamics and the geometry of the explosion. While the existence of this instability has been firmly established, its physical origin remains somewhat uncertain. Two mechanisms have indeed been proposed to explain its linear growth. The first is an advective-acoustic cycle, where the instability results from the interplay between advected perturbations (entropy and vorticity) and an acoustic wave. The second mechanism is purely acoustic and assumes that the shock is able to amplify trapped acoustic waves. Several arguments favouring the advective-acoustic cycle have already been proposed, however none was entirely conclusive for realistic flow parameters. In this article we give two new arguments which unambiguously show that the instability is not purely acoustic, and should be attributed to the advective-acoustic cycle. First, we extract a radial propagation timescale by comparing the frequencies of several unstable harmonics that differ only by their radial structure. The extracted time matches the advective-acoustic time but strongly disagrees with a purely acoustic interpretation. Second, we present a method to compute purely acoustic modes, by artificially removing advected perturbations below the shock. All these purely acoustic modes are found to be stable, showing that the advected wave is essential to the instability mechanism. 
\end{abstract}

\begin{keywords}
instabilities --- waves --- shock waves --- supernovae: general --- methods: analytical
\end{keywords}

\section{Introduction}

Several observational as well as theoretical arguments show that most core collapse supernovae are asymmetric explosions. The observational evidence include polarisation measures \citep{leonard06}, double peaked oxygen lines \citep{maeda08}, and the large velocities of neutron stars \citep{lai01}. On the theoretical side, the failure of realistic spherically symmetric calculations to produce an explosion \citep{liebendorfer01} suggests that the breaking of the initial spherical symmetry is essential to the explosion mechanism. Indeed most proposed mechanisms are fundamentally asymmetric: the modern version of the neutrino-driven mechanism \citep{marek09}, the acoustic mechanism \citep{burrows06}, and the magneto-centrifugal mechanism \citep{wheeler02}.

The spherical symmetry can be broken by several phenomena. In the magneto-centrifugal mechanism, the asymmetry is due to a fast rotation in conjunction with a strong magnetic field. In the case of moderate rotation, two non radial instabilities can develop during the stalled shock phase: the neutrino-driven convection, and the stationary accretion shock instability usually called SASI which induces non spherical oscillations of the shock \citep{blondin03}. Of these two instabilities, SASI is the most efficient in creating a global asymmetry \citep{foglizzo06} and helping the explosion \citep{marek09,burrows06}. In addition to its crucial role in the explosion mechanism, SASI has a number of potentially important consequences: it affects the kick and the spin of the neutron star \citep{scheck06,blondin07a}, causes an emission of gravitational waves \citep{kotake07}, and creates a time dependence of the neutrino signal \citep{marek09b}.

But what is SASI ? Which physical phenomenon drives these shock oscillations ? Surprisingly, the answer to this basic question remains somewhat uncertain. Indeed, two mechanisms have been proposed -- the advective-acoustic cycle \citep{foglizzo07} and a purely acoustic mechanism \citep{blondin06} -- and distinguishing them remains a controversial issue. The aim of this article is to clarify this question.

 In Section~\ref{sec:review}, we review previous works by explaining the two instability mechanisms and the arguments proposed to distinguish them. On the basis of these works, the advective-acoustic cycle is generally favoured but the purely acoustic mechanism cannot be ruled out for realistic flow parameters. In Section~\ref{sec:radial_time}, we propose a new method to extract a radial propagation time from the frequency of the unstable modes. For this purpose we compare the frequencies of several harmonics that differ only by their radial structure. As the radial advective-acoustic and purely acoustic times are significantly different, this allows us to distinguish between the two cycles and discards the purely acoustic mechanism. In Section~\ref{sec:acoustic_modes},  we describe a method to compute purely acoustic modes, which are shown to be stable. In Section~\ref{sec:details}, we give a detailed analysis of the frequency spectra of advective-acoustic and purely acoustic cycles, which confirms the validity of the method used in Section~\ref{sec:radial_time}. Finally in Section~\ref{sec:conclusion}, we summarise our work and conclude that the instability mechanism behind SASI is the advective-acoustic cycle.

\section{The two proposed mechanisms: summary of previous works}
	\label{sec:review}

\subsection{Different models of SASI}
	\label{sec:SASI_models}
	
SASI has been observed to grow in many different models of stellar collapse, which have very diverse degrees of complexity. These include state of the art realistic numerical simulations \citep{marek09,burrows06}, as well as models where successive simplifications have been made: approximate neutrino transport \citep{scheck08}, neglect of the stellar structure \citep{ohnishi06,yamasaki07,iwakami08}, simple equation of state with the effect of neutrinos parameterised by cooling/heating functions \citep{blondin03,blondin06,foglizzo07,yamasaki08,fernandez09a,fernandez09b}, adiabatic approximation \citep{blondin07a}, and finally the most simplified of all these models is the planar toy model \citep{foglizzo09,sato09}. Very often, the simpler the model, the deeper the physical understanding. For example, models neglecting the heating have clearly shown that SASI is distinct from neutrino-driven convection \citep{blondin03}, a fact missed by most realistic models. The neglect of the stellar structure allows to set up an initial steady state, which can be studied with a perturbative analysis \citep{foglizzo07,yamasaki07}. Finally, the simplest of these models, the planar toy model, allows for an analytical treatment where the instability mechanism can be clearly demonstrated. 

It is widely accepted that the same instability is at work in all these models. Therefore, although we base our investigation on the model of \citet{blondin06}, we expect our conclusions to hold more generally. We choose this model because it is simple enough to perform a perturbative analysis with clean boundary conditions. This idealised flow was first described by \citet{houck92} and later studied by \citet{blondin06} to describe SASI. It has then become a classical model of SASI and has been studied also by \citet{foglizzo07,fernandez09a,fernandez10,guilet10b}. In this flow, the fluid is modelled as a non self-gravitating perfect gas with an adiabatic index of 4/3. The effects of neutrinos are simply modelled by a cooling function, while neutrino heating is neglected. The proto-neutron star is treated as a hard surface on which the cooling fluid settles down. The shock is stationary at a radius which can be freely chosen by adjusting the normalisation of the cooling function. This model and its linear analysis will be extensively used in the remainder of the paper. We refer the reader to \cite{foglizzo07} for a description of the linear eigenmodes calculation, and to Appendix~\ref{sec:equations} for a slightly different (but equivalent) formulation of the equations governing it, which has been used in this paper.

\subsection{The two mechanisms}
	\label{sec:review_mechanisms}

\begin{figure*}
\centering
\includegraphics[width=2\columnwidth]{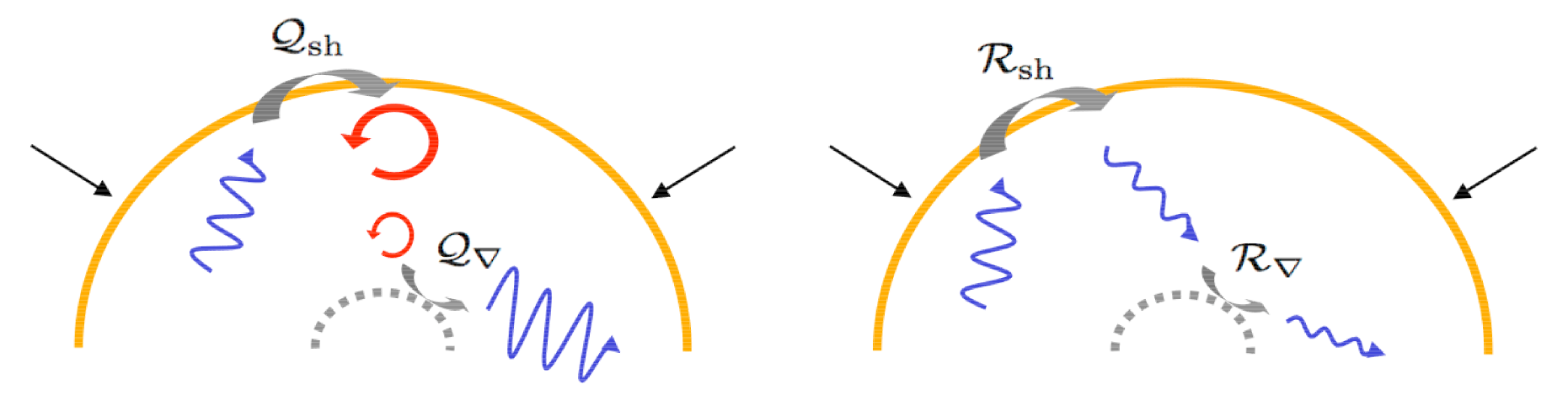}
	\caption{Schematic representation of the linear instability mechanism proposed to explain SASI (adapted from \citet{foglizzo09b}). The advective-acoustic cycle (left panel) comprises two coupling processes between an advected wave (represented here by the circular red arrows) and an upward propagating acoustic wave (represented by wavy arrows). These coupling processes take place at the shock and in a region of strong deceleration close to the surface of the proto-neutron star. Note that while advection is purely radial, acoustic propagation can have a significant transverse component. In the purely acoustic mechanism a trapped acoustic wave is amplified through its interaction with the shock. Formally, it can be represented by two coupling processes between two acoustic waves, propagating up and down (right panel, note the concurrent transverse propagation). The advective-acoustic and purely acoustic cycles have been represented respectively unstable and stable, in anticipation of the conclusion of this article.}
	\label{fig:lineaire:mecanismes}
\end{figure*}

The two mechanisms proposed to explain the growth of SASI are schematically illustrated by Figure~\ref{fig:lineaire:mecanismes}. The advective-acoustic cycle (left panel) is a cycle between two waves : an advected wave composed of entropy and vorticity and an acoustic wave propagating outward \citep{foglizzo00,foglizzo02,blondin03,foglizzo07}. Note that the acoustic wave also propagates in the transverse direction. These waves are coupled via two coupling processes taking place at the shock and at a smaller radius in the deceleration region. The acoustic wave reaching the shock makes it oscillate and creates the entropy-vorticity wave. This wave is advected towards the proto-neutron star (hereafter called PNS) and its deceleration creates an acoustic feedback thus closing the loop. The cycle is unstable if the amplitude of the wave at the end of a cycle is larger than it was at the beginning. 

The purely acoustic mechanism has been described in \cite{blondin06} as an acoustic wave trapped in the postshock cavity. This acoustic mode is assumed to be ampliÞed through its interaction with the shock. Even when the acoustic propagation is mostly azimuthal some radial propagation is also present between the shock wave and the inner turning point (Fig. 6 of \cite{blondin06}). Such an acoustic path can be interpreted as a purely acoustic cycle between the shock and the turning point (Figure~\ref{fig:lineaire:mecanismes}, right panel). We stress that this description is quite general and does not assume a purely radial propagation. The transverse propagation is not restricted or disregarded neither here nor anywhere in this paper.
 
Any of these two cycles could in principle be unstable depending on the efficiency of the different coupling processes. For example, the purely acoustic cycle might be unstable if the acoustic wave is amplified when it is reflected at the shock, as assumed by \citet{blondin06}. Alternatively, the advective-acoustic cycle can be unstable if gradients below the shock allow for an efficient acoustic feedback, and if the shock is strong enough to efficiently create an entropy-vorticity wave \citep{foglizzo00,foglizzo01,foglizzo02,foglizzo09}. By studying a simplified model of the advective-acoustic cycle, \citet{foglizzo09} could explain salient features of SASI, in particular that it is dominated by low frequency and large scale non-radial modes. On the other hand, the purely acoustic cycle lacks a reference model where the properties of such acoustic instability could be unambiguously determined. As a consequence, it has not so far explained the basic features of SASI.

\subsection{Timescales}
	\label{sec:timescales}

A large part of this article deals with timescales: in particular how they determine the oscillation frequency, and how this can be used to discriminate between the two mechanisms. It is therefore useful to recall the characteristic timescales of the two cycles. The duration of a radial advective acoustic cycle can be written as:
\begin{equation}
\tau_{\rm aac} \equiv \int_{r_{\rm coup}}^{r_\sh}\frac{1}{1-\M}\frac{\dd r}{|v|}, 
	\label{eq:taac_spherical}
\end{equation}
where $r_{\rm coup}$ is the effective coupling radius where the acoustic feedback is produced\footnote{Note that this is a simplified description of the instability: in reality the feedback is distributed radially rather than produced at a single radius.}. \citet{foglizzo07} and \citet{scheck08} have suggested that the coupling radius is close to the location where the deceleration is strongest, however there remains some uncertainty in the precise location of this coupling. Similarly, the duration of a radial purely acoustic cycle is:
\begin{equation}
\tau_{{\rm ac}r} \equiv \int_{r_{\rm ref}}^{r_\sh}\frac{2}{1-\M^2}\frac{\dd r}{c}, 
	\label{eq:tacr_spherical}
\end{equation}
where $r_{\rm ref}$ is the radius where the acoustic wave propagating down is reflected (i.e. its turning point). Finally, another acoustic timescale which plays a role potentially for both of the cycles is the azimuthal acoustic time:
\begin{equation}
\tau_{{\rm ac}\phi}(r) \equiv \frac{2\pi r}{c}.
	 \label{eq:tacphi}
\end{equation}
The two radial timescales suffer some uncertainty due to the lack of knowledge about $r_{\rm coup}$ and $r_{\rm ref}$, both of which can in principle depend non trivially on the mode considered. In the model studied in this paper and briefly described in Section~\ref{sec:SASI_models}, it is however possible to obtain a useful upper limit by supposing that these two radii coincide with the PNS surface\footnote{With the cooling function considered here ($\alpha=3/2$, $\beta=5/2$), the advection time until the PNS surface is finite. However with a less violent cooling law ($\alpha=1$, $\beta=6$) this time is infinite, so that no useful upper limit can be obtained. It is then more relevant to estimate the coupling radius to be where the velocity gradient is strongest \citep{foglizzo07}.}. In the following, we choose this prescription, keeping in mind that a shorter time (hence a higher frequency) can be obtained by changing $r_{\rm coup}$ and $r_{\rm ref}$. The azimuthal acoustic time also suffers an uncertainty due to the radius at which it should be estimated. Here again a useful upper limit can be obtained by choosing the shock radius $r_\sh$.

\subsection{WKB description of the two cycles (following \citet{foglizzo07})}
	\label{sec:review_WKB}
In order to quantitatively describe the two cycles, one needs to separate perturbations below the shock into the three types of waves: two acoustic waves (propagating up and down) and the advected wave. Below the shock the flow is close to adiabatic (\citet{foglizzo07}, and see Section~\ref{sec:acoustic_modes}), thus the entropy and vorticity are simply advected. The main difficulty comes from the acoustic perturbations, which can be separated between waves propagating up and down only in the WKB approximation. The WKB approximation is valid if the wavelengths of the acoustic and advected waves are small compared to the length scale of the gradients.  It can therefore be applied only to the modes with a high enough frequency, i.e. satisfying:
\begin{equation}
\omega \frac{r_\sh}{c_\sh} \gg 1.    \label{eq:WKB_criterion}
\end{equation}
\citet{foglizzo07} performed such a WKB analysis of the flow considered by \citet{blondin06}. In the following, we summarise the principle of this analysis, which could in principle be applied more generally to other flows.

Constants describing the coupling efficiency can be defined in the following way. When an acoustic wave propagating up with an amplitude $\delta f^{-}$ hits the shock, it creates an advected wave with an amplitude $\delta f^{\rm adv}$ as well as an acoustic wave propagating down with an amplitude $\delta f^{+}$ \footnote{the definition of the variable $\delta f$ is recalled in Appendix~\ref{sec:equations}. Choosing another variable to measure the amplitude would change the coupling constants at the shock and in the gradients, but not the global cycle constant determining the stability of the cycles.}. The ratio of the amplitude of the waves created to that of the incoming wave defines the coupling constants: $Q_\sh \equiv \delta f^{\rm adv}/\delta f^-$ and $R_\sh \equiv \delta f^{+}/\delta f^-$. These can be computed analytically using the boundary conditions at the shock and by separating the perturbations into the three types of waves owing to the WKB method \citep{foglizzo05,foglizzo09}. 

When an entropy-vorticity wave is advected towards the PNS, it creates an acoustic wave propagating up with an efficiency characterised by the ratio of the amplitudes (measured at the shock): $Q_\nabla \equiv \delta f^{-}/\delta f^{\rm adv}$. Similarly the reflection in the gradients of the acoustic wave propagating down is characterised by: $R_\nabla \equiv \delta f^{-}/\delta f^{+}$. These two constants can be computed numerically by integrating the differential system describing the perturbations, imposing modified boundary conditions at the shock ($\delta f^+=0$ to obtain $Q_\nabla$, $\delta f^{\rm adv} = 0$ to obtain $R_\nabla$). 

Finally, using these four coupling constants, one can define the global constants that describe the efficiency of the two cycles (the amplitude ratio between the beginning and the end of the cycles): $Q \equiv Q_\sh Q_\nabla$ and $R \equiv R_\sh R_\nabla$. The cycle constants are complex numbers that depend on the complex frequency $\omega$. Since the two cycles take place simultaneously, an eigenmode must satisfy the following relation \citep{foglizzo00}:
\begin{equation}
Q + R = 1.
	\label{eq:QR}
\end{equation}

If these constants are calculated with a real frequency then the instability criterion for a single cycle can be written: $|Q|>1$ or $|R|>1$ (depending on the cycle considered), which is equivalent to the criterion stated in Section~\ref{sec:review_mechanisms}. The growth rate can then be approximated by:
\begin{equation}
\omega_i \simeq \frac{\log|Q|}{\tau},
\end{equation}
where $\tau$ is the duration of the cycle considered (here the advective-acoustic one). When a second cycle is present, it can interfere constructively or destructively with the first cycle depending on their relative phase. This causes oscillations of the growth rate, which are indeed observed in the eigenspectrum of SASI (Figure~7 of \citet{foglizzo07}). These oscillations contain information about the efficiency and the timescale of the two cycles at work. \citet{foglizzo07} could thus extract directly from the eigenspectrum the values of the constants $Q$ and $R$ (note that this is independent from the WKB analysis, and only assumes the presence of two cycles). The value of the cycle constants obtained from the WKB method and from the analysis of the eigenspectrum are in very good agreement (their Figures~8 and 9). They show that the advective-acoustic can be unstable with a cycle constant reaching $Q \sim 6$, while the purely acoustic cycle is always stable with typically $R\sim0.5$.

\subsection{Which cycle drives SASI?}
	\label{sec:review_arguments}
The WKB analysis of \citet{foglizzo07} summarised in the last subsection has proven that the advective-acoustic cycle is unstable and can explain SASI, while the purely acoustic cycle is stable. However this result is valid only for high frequency modes, satisfying Equation~\ref{eq:WKB_criterion}. The first few harmonics do not satisfy this inequality as $\omega r_\sh/c_\sh \sim 1$ for the fundamental mode. One may consider that the WKB method can be safely applied for the tenth harmonics and higher (satisfying $\omega r_\sh/c_\sh > 10$). Unfortunately these high frequency modes are unstable only for large shock radii $r_\sh/r_*>5-10$ (depending on the cooling function chosen). Thus strictly speaking in the more realistic range of shock radius $2< r_\sh/r_*<5$, the WKB analysis cannot conclude on the instability mechanism. \citet{foglizzo07} have however argued by a continuity argument that it is not necessary to invoke a different instability mechanism for the low frequency modes. Indeed, the eigenspectrum vary smoothly both with the shock radius and mode frequency, and the eigenfunction of the low frequency modes resembles that of the high frequency modes. It may thus seem natural to assume that all SASI modes are due to the advective-acoustic cycle. Another argument in this direction is that the WKB approximation usually gives quite good results even when the small parameter (here $c_\sh/(r_\sh\omega)$) is of order unity. This is however not a formal proof, and it cannot be excluded that the low frequency modes at moderate shock radius originate from a different instability mechanism.

\citet{laming07} has studied analytically the stability of a spherical accretion shock by establishing an approximate dispersion relation, using a method inspired from \citet{vishniac89}. This equation contains terms due to advection, which \citet{laming07} proposes to remove in order to assess the importance of the advective-acoustic cycle in the instability mechanism. The conclusion he initially reached by this mean depended on the radius of the shock. At large shock radii, the omission of the advective term stabilised the flow, thus favouring the advective-acoustic cycle. On the contrary, at small shock radii advection affected very little the growth rate, favouring a purely acoustic interpretation of the instability. However, this analysis has been criticised by \citet{yamasaki08} because the dispersion relation was established with the use of dubious approximations. Furthermore an erratum published later on \citep{laming08} corrects a few mistakes and changes the conclusion: the advective-acoustic cycle is then favoured for all shock radii. 

A number of authors have tried to distinguish the two mechanisms by using the oscillation frequency observed in the simulations. These studies have led to diverging conclusions: \citet{ohnishi06} concluded that the advective-acoustic cycle is responsible for the instability, while \citet{blondin06} concluded in the favour of the purely acoustic mechanism. This confusion probably comes from the fact that both of the mechanisms can explain the oscillation frequency if one makes the right assumptions. Indeed, \citet{ohnishi06} excluded the purely acoustic mechanism because the oscillation period is much longer than the \emph{radial} acoustic time, while \citet{blondin06}  have shown that the oscillation period could be explained if one considers a non radial acoustic path. On the other hand, \citet{blondin06} excluded the advective-acoustic cycle without considering the possibility that the acoustic feedback be created at a larger radius than the PNS surface. The determination of the frequency of each cycle thus suffers some uncertainty due to our lack of knowledge of the coupling radius (for the advective-acoustic cycle), and of the non-radial path to be considered (for a purely acoustic cycle). Later, \citet{scheck08} have shown that the oscillation period observed in their simulations was compatible with both mechanisms if the advective-acoustic coupling radius lies where the deceleration is maximum, or if an \emph{ad hoc} non radial acoustic path is chosen (their Figure~16). 

\citet{fernandez09a} have proposed an argument that can be applied to the regime in which the WKB method is inapplicable. They followed the variation of a SASI mode growth rate as the shock radius is varied, which changes the ratio of the azimuthal acoustic time to the advection time from the shock to the PNS. They observed that the mode has a maximum growth rate when the two times coincide, and interpreted this as an indication that the advective-acoustic cycle is at work (because both advection and acoustic times play a role in the instability). While this observation is indeed suggestive that the advection plays a role in determining the growth rate, it does not rule out the possibility of the acoustic cycle playing a dominant role with only a minor help from the advection. 

Some more indirect information about the instability mechanism can be found by considering the effect of a magnetic field on SASI. \citet{endeve10} have performed numerical simulations of SASI in the presence of a radial magnetic field (a split monopole). They obtained the surprising result that a magnetic field, which magnetic pressure is negligible compared the thermal one ($P_{\rm mag}/P_{\rm th} < 10^{-3}$), is capable of entirely stabilising SASI (in their model 2DB14Am). This seems to contradict a purely acoustic mechanism because the acoustic waves should be almost unaffected by such a weak magnetic field. By contrast, this result can be naturally interpreted in the framework of the advective-acoustic cycle. Indeed \citet{guilet10a} have shown that the advective-acoustic cycle is significantly affected by a magnetic field if the Alfv\'en speed is comparable to the advection velocity, which can happen even if the magnetic pressure is negligibly small. In their model 2DB14Am, the Alfv\'en speed is indeed comparable to the advection velocity in the vicinity of the PNS.

To summarise this subsection, the advective-acoustic mechanism is favoured by most arguments. However none of these arguments can rule out the purely acoustic mechanism for realistic flow parameters. The goal of this paper is to find a method to definitely distinguish the two mechanisms for realistic flow parameters.

\subsection{This article}

\begin{figure}
\includegraphics[width=\columnwidth]{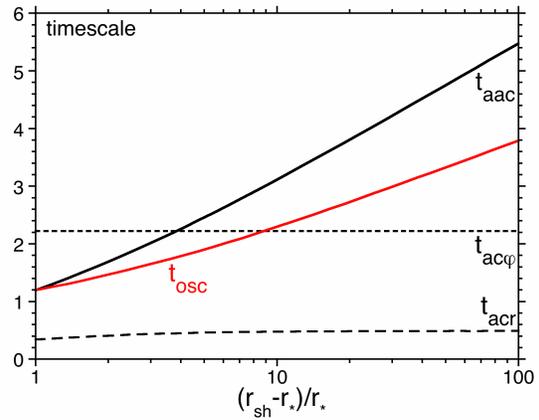}
	\caption{Comparison of the oscillation period of the fundamental $l=1$ SASI mode with the advective-acoustic and purely acoustic times. This calculation is made in the classical model of SASI briefly described in Section~\ref{sec:SASI_models}, with the parametters $\alpha=3/2$, $\beta=5/2$, $\M_1=\infty$ (as in the rest of the paper). All timescales are in units of $r_\sh/v_\sh$. The oscillation period ($t_{\rm osc}$, red line) is comparable both to the radial advective-acoustic time ($t_{\rm aac}$, full black line, Equation~(\ref{eq:taac_spherical})) and to the azimuthal acoustic time at the shock ($t_{\rm ac\phi}$, short dashed line, Equation~(\ref{eq:tacphi})). By contrast, the radial acoustic time is much shorter than these timescales ($t_{\rm acr}$, long dashed line, Equation~(\ref{eq:tacr_spherical})). For all timescales an upper limit is shown as discussed in Section~\ref{sec:timescales} ($r_{\rm coup}=r_*$, $r_{\rm ref}=r_*$, and $\tau_{{\rm ac}\phi}$ is estimated at the shock radius).}
	\label{fig:period_SASI}
\end{figure}

As discussed in the last subsection, previous studies were not successful in determining the instability mechanism from SASI oscillation frequency, because the fundamental mode frequency can be explained by both mechanisms. This is due to the fact, that the (radial) advective-acoustic time and the \emph{azimuthal} acoustic time are both comparable to the oscillation period, as is illustrated  by Figure~\ref{fig:period_SASI}. In Section~\ref{sec:radial_time} we present a method to avoid this difficulty by extracting a \emph{radial} time from the frequencies of several unstable radial harmonics. The large difference between acoustic and advective-acoustic radial times allows us to distinguish clearly the two mechanisms (the radial acoustic time is shorter by a factor 3 to 10, as shown in Figure~\ref{fig:period_SASI}).

Another reason for the remaining ambiguity on the instability mechanism is that, for realistic flow parameters, the unstable modes cannot be described by the WKB method. In Section~\ref{sec:acoustic_modes} we present a new method to compute purely acoustic modes, which partially avoids this difficulty. This method makes use of the fact that the wavelength of advected perturbations is shorter than the acoustic wavelength. As a consequence, the identification of advected perturbations through a WKB approximation is valid at lower frequency.

Finally, in Section~\ref{sec:details} we develop an analytical description of the frequency spectrum of the two cycles, in order to confirm the validity of the argument presented in Section~\ref{sec:radial_time}. This analytical treatment is checked against the acoustic modes computed in Section~\ref{sec:acoustic_modes}.

In the following we apply our two arguments to the classical model of SASI briefly described in Section~\ref{sec:SASI_models}. The reader is referred to \citet{blondin06} and \citet{foglizzo07} for a more detailed description. We focus on a shock radius $r_\sh = 2.5 r_* $, which is typical of supernovae and for which the unstable modes cannot be described by the WKB method. We also choose the following parameters: $\alpha=3/2$, $\beta=5/2$, $\M_1=\infty$.

\section{Extraction of a radial propagation timescale from the eigenspectrum}
	\label{sec:radial_time}
Previous studies that tried to determine the instability mechanism using the oscillation frequency \citep{blondin06,ohnishi06,scheck08} have based their discussion entirely on the mode that grew fastest in their simulation. This most unstable mode is usually the lowest frequency $l=1$ mode, and its frequency is consistent with a radial advective-acoustic time as well as with a non-radial acoustic time. The specificity of our study is that we consider other (less) unstable modes, which can provide us with additional constraints on the instability mechanism. These modes are accurately computed by a linear analysis \citep{foglizzo07}, but may be hard to distinguish clearly in numerical simulations (although they should be present) because they grow more slowly and thus have a smaller amplitude than the most unstable mode
. We focus in particular on higher frequency harmonics having the same transverse structure as the most unstable $l=1$ mode. For most astrophysically relevant values of the shock radius, several of these modes are unstable. For example, for our fiducial value $r_\sh=2.5r_*$, three $l=1$ modes are unstable (Figure~\ref{fig:SASI_spectrum}). We emphasise that these modes have a higher frequency than the fundamental mode, but not high enough to allow for a WKB analysis.

\begin{figure*}
\centering
\includegraphics[width=2\columnwidth]{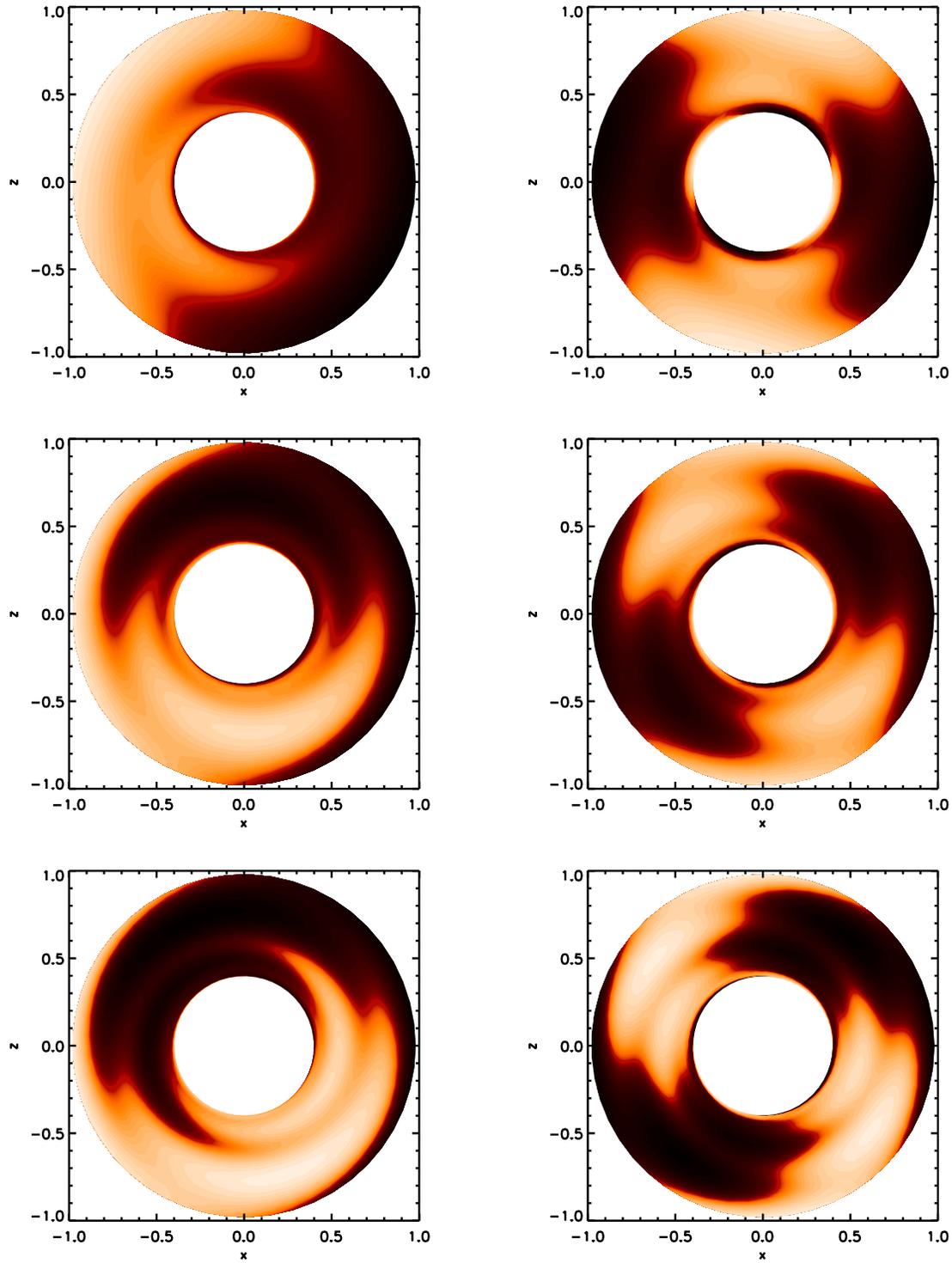}
	\caption{Normalised pressure perturbations $\delta P/P$ in the equatorial plane associated to different SASI modes. $l=1$, $m=1$ modes are depicted in the left column, while  $l=2$, $m=2$ are in the right column. The three panels in each column correspond to the three lowest frequency modes for the given spherical harmonics index (low to high frequency represented from top to bottom). This illustrates how modes with the same spherical harmonics index differ by their radial structure while they share their transverse structure: higher frequency harmonics have a more complex radial structure.}
	\label{fig:eigenfunctions}
\end{figure*}

What timescale determines the frequency change of the higher harmonics as compared to the fundamental mode ? We argue that this timescale cannot correspond to an azimuthal propagation because all these modes share the same transverse structure (determined by the spherical harmonics index $l$). As illustrated by Figure~\ref{fig:eigenfunctions}, these modes differ only by their radial structure suggesting that the frequency change should instead be controlled by a radial time, either advective-acoustic or purely acoustic. As a consequence, we suggest that a radial propagation timescale can be extracted from the frequency difference between two successive modes that share the same transverse structure (i.e. same spherical harmonics). We define such a time in the following way:
\begin{equation}
t_{n-n+1} \equiv \frac{2\pi}{\omega_r(n+1) - \omega_r(n)}
	\label{eq:radial_time_extraction}
\end{equation}
where $\omega_r(n)$ and $\omega_r(n+1)$ are the frequencies of the n-th and n+1-th mode of SASI for a given spherical harmonics index $l$ (ranking the modes from the lowest to the highest frequency). From the physical argument stated above, we expect this time to be of the order of the radial acoustic time if the instability mechanism were purely acoustic, or of the order of the radial advective-acoustic time if SASI were due to the advective-acoustic cycle. Given the qualitative nature of the argument, we may not however expect a precise agreement. Indeed the analysis of the frequency spectrum detailed in Section~\ref{sec:details} confirms that this is only approximately true. However thanks to the large difference between radial times (Figure~\ref{fig:period_SASI}), this approximate estimate is enough to be conclusive. We note that this method bears some similarity with the extraction of the radial structure of the Sun using the "large frequency spacing" in helioseismology \citep{deubner84}.

\begin{figure}
\centering
\includegraphics[width=\columnwidth]{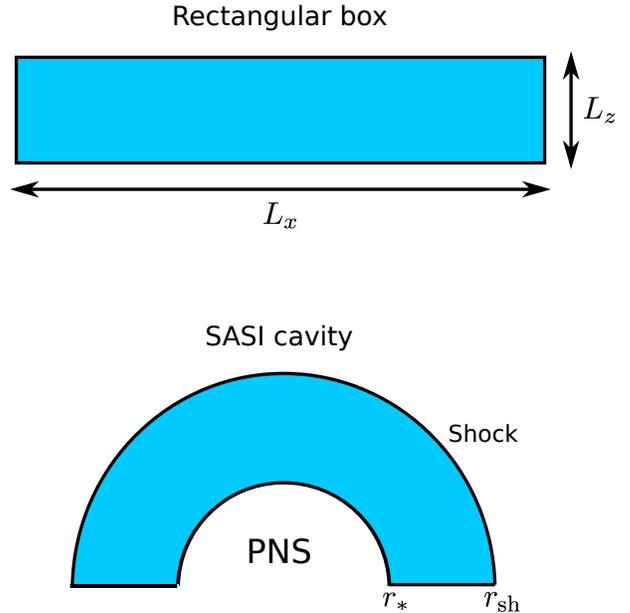}
	\caption{Schematic illustration of the analogy between acoustic modes in a rectangular box and in stellar core collapse. }
	\label{fig:box_analogy}
\end{figure}

Some light can be shed on the above argument by drawing an analogy with the very classical case of acoustic modes in a rectangular box (illustrated by Figure~\ref{fig:box_analogy}). Let us assume that the box is filled with a uniform gas at rest and delimited by reflecting rigid walls. The frequency eigenspectrum of acoustic modes is then:
\begin{eqnarray}
\omega &=& 2\pi \left\lbrack\frac{n_z^2}{\tau_{{\rm ac}z}^2} + \frac{n_x^2}{\tau_{{\rm ac}x}^2}\right\rbrack^{1/2},
	\label{eq:freq_box} \\
	&=& 2\pi \frac{n_z}{\tau_{{\rm ac}z}}\left\lbrack 1 + \frac{n_x^2\tau_{{\rm ac}z}^2}{n_z^2\tau_{{\rm ac}x}^2}  \right\rbrack^{1/2}.
\end{eqnarray}
where $\tau_{{\rm ac}x} \equiv 2L_x/c$ and $\tau_{{\rm ac}z} \equiv 2L_z/c$ are the back and forth acoustic propagation times in the $x$ and $z$ directions, while $n_x$ and $n_z$ are the number of half wavelengths in the width of the box (in $x$ and $z$ direction). To draw the analogy, we consider a box elongated in the $x$-direction, with $z$ representing the radial direction and $x$ representing the transverse direction ($\theta$ or $\phi$). The correspondences are thus $L_x \rightarrow \pi r$, $L_z \rightarrow r_\sh - r_*$ and $n_x \rightarrow l$. $n_z$ distinguishes the different modes that share the same spherical harmonics index. $n_z=0$ corresponds to the lowest frequency fundamental mode, with a frequency $\omega=2\pi n_x/\tau_{{\rm ac}x}$ that is dictated by the transverse acoustic time. Higher integer values of $n_z$ correspond to the higher frequency modes, which are the focus of this section. As $n_z$ is increased the frequency increases due to the first term in Equation~(\ref{eq:freq_box}), which contains the acoustic time $\tau_{{\rm ac}z} $ in the $z$-direction (analogous to the radial acoustic time). This confirms the qualitative argument given in the last paragraph. As the frequency is not simply proportional to $2\pi n_z/\tau_{{\rm ac}z}$ , the frequency difference between two successive modes is not exactly the "radial" frequency $2\pi/\tau_{{\rm ac}z}$, suggesting that the Equation~(\ref{eq:radial_time_extraction}) is only an approximate estimate even in this simplest case. However the extracted timescale does tend toward the radial time in the limit where $n_x^2/\tau_{{\rm ac}x}^2 \ll n_z^2/\tau_{{\rm ac}z}^2$. As the azimuthal acoustic time is significantly larger the radial time, we can expect the method of extraction of a radial timescale to give reasonable results.  
 
\begin{figure}
\centering
\includegraphics[width=\columnwidth]{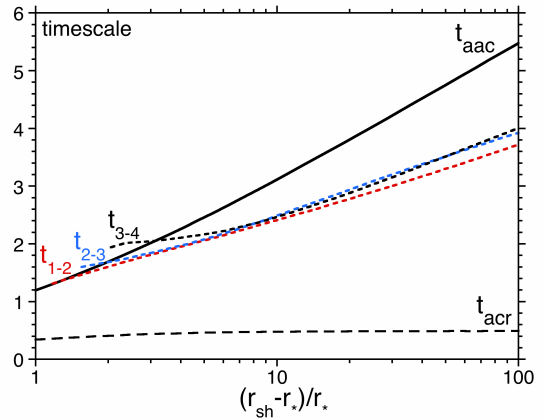}
	\caption{Extraction of a radial propagation time from the unstable $l=1$ SASI modes (Equation~(\ref{eq:radial_time_extraction}), short dashed lines) and comparison with the advective-acoustic time (Equation~(\ref{eq:taac_spherical}), full line) and the purely acoustic radial time (Equation~(\ref{eq:tacr_spherical}), long dashed line). The times extracted from the eigenspectrum are all much longer than the acoustic time (by a factor 3.5 to 7), which rules out a purely acoustic origin of these unstable modes. By contrast, they are comparable to the advective-acoustic time, thus confirming the advective-acoustic cycle as the origin of the instability. All timescales are in units of $r_\sh/v_\sh$. }
	\label{fig:radial_timescale}
\end{figure}

The radial timescales extracted by using the mode duplets $n=1-2$, $n=2-3$, and $n=3-4$ (all of them with $l=1$) are compared to the advective-acoustic and purely acoustic radial times in Figure~\ref{fig:radial_timescale}. The timescales extracted from mode duplets are represented only on the radius interval in which both of the modes are unstable. The three extracted timescales are close to each other, which gives some confidence on their physical relevance. Furthermore, they roughly match the advective-acoustic time: within $\sim 15\%$ for moderate shock radius in the range $r_\sh=2-5r_*$ (the focus of this article), and within $\sim 50\%$ for larger shock radii. By contrast, the extracted timescales are significantly longer than the radial acoustic time: the ratio varies between 3.5 for small shock radii and 6 for larger shock radii. This result clearly excludes the purely acoustic mechanism and argues in favour of the advective-acoustic cycle as the origin of SASI.

The difference at large shock radii between the advective-acoustic and the extracted timescale does not prevent the argument to be conclusive for several reasons. First, as compared to the large discrepancy with the purely acoustic time, this disagreement is quite slight. Second, it can easily be interpreted in the context of an advective-acoustic cycle by adjusting the effective coupling radius. Indeed the advective-acoustic time has been computed between the shock and the PNS surface, and as such should be considered as an upper limit. By contrast, it is impossible to reconcile in the same way the acoustic time with the measured timescales, because the acoustic time is too \emph{short}. Finally we note that the instability mechanism at large shock radii was anyway not much debated, because the WKB analysis of \citet{foglizzo07} can be conclusively applied to higher frequency harmonics.

\section{Computation of purely acoustic eigenmodes}
	\label{sec:acoustic_modes}

\begin{figure*}
\centering
\includegraphics[width=\columnwidth]{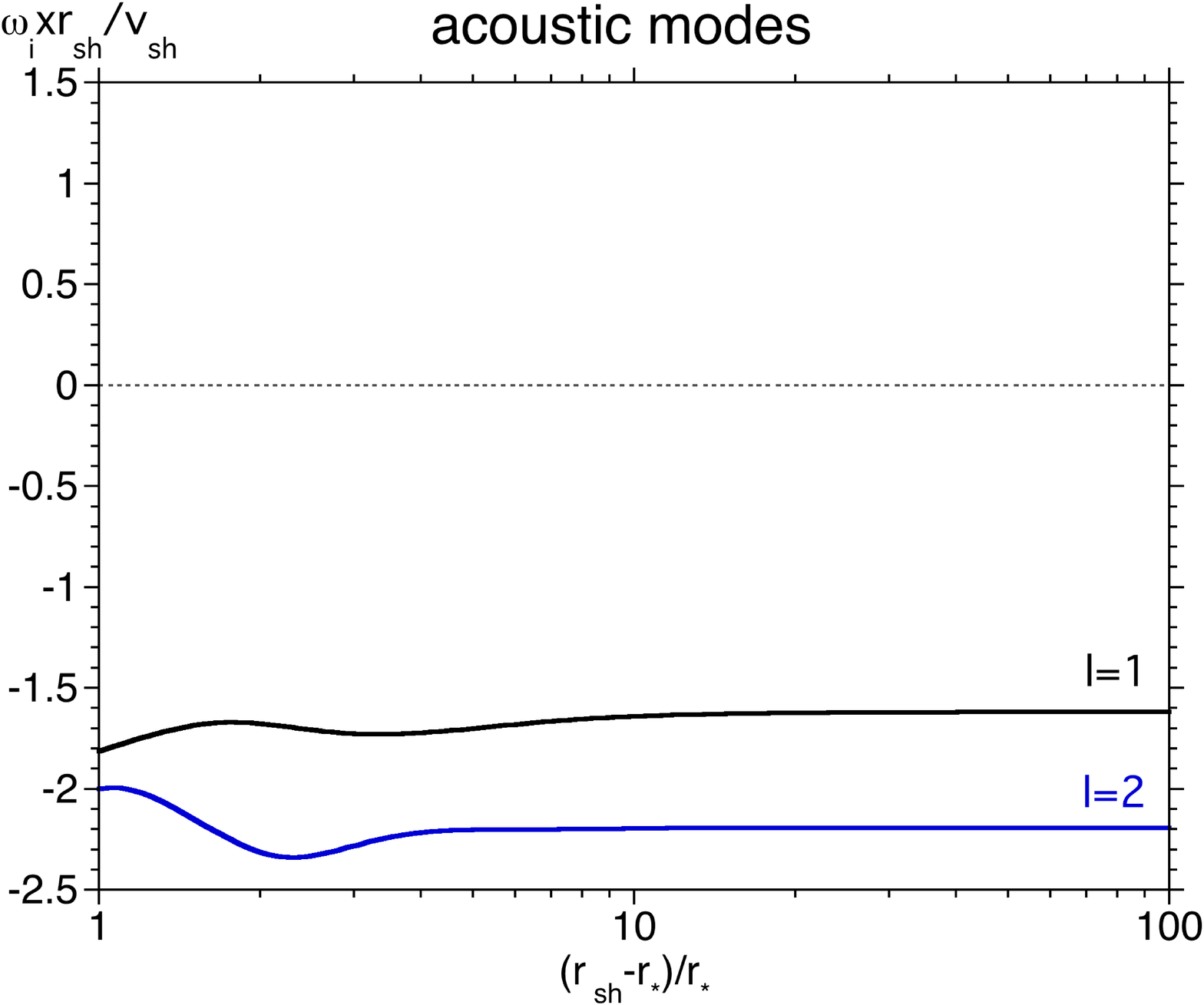}
\includegraphics[width=\columnwidth]{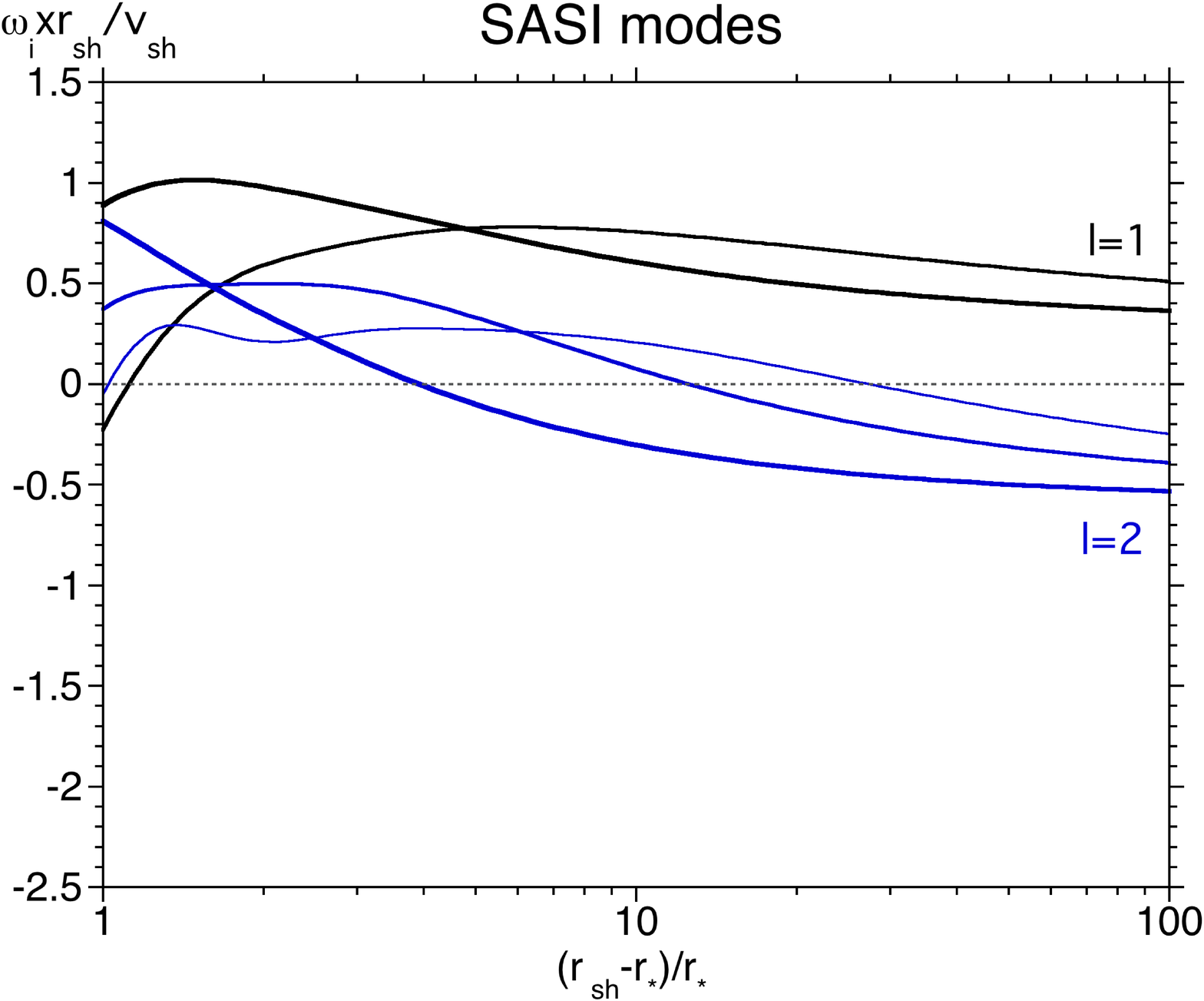}
	\caption{Growth rate of the purely acoustic modes (left panel) compared with the growth rate of SASI (right panel). $l=1$ modes are shown in black and $l=2$ modes in blue. For each spherical harmonics index, the most unstable mode is plotted (in the case of SASI, several harmonics are shown with thiner line for higher frequency modes).}
	\label{fig:acoustic_growth_rate}
\end{figure*}

In this section, a method to compute purely acoustic modes is described. In this method, one separates the advected perturbations just below the shock, and subtract them from the total perturbations at the shock. Let us emphasise that a WKB description of the acoustic waves is not needed, since we do not need to distinguish acoustic waves propagating up and down. As a result, the method is applicable to lower frequency modes. We stress that we do not make any assumption regarding the transverse propagation of acoustic waves. The WKB method is applicable when the wavelength of acoustic waves is shorter than the scale of the gradients: $\omega r_\sh/c_\sh \gg 1$, while the separation of advected perturbations only requires that the advection wavelength be shorter than the gradient scale: 
\begin{equation}
\omega r_\sh/v_\sh \gg 1.
	\label{eq:WKBadv_criterion}
\end{equation}
For example for $r_\sh=2.5r_*$, the first three $l=1$ SASI modes
 have $\omega r_\sh/v_\sh \sim 4.5-9-13$. We may thus expect that the advected wave can be extracted reasonably well for the second and third modes, but only marginally for the first mode.

Another assumption implicitly made here is that, just below the shock, the entropy is simply advected, which is true if the flow is adiabatic. The flow just below the shock is indeed usually rather close to adiabaticity. The quasi-adiabaticity of the postshock flow can be quantified by the dimensionless parameter $r\p S/\p r(r_\sh)$ (where $S$ is a dimensionless entropy defined in \citet{foglizzo07}), which ranges from $7.10^{-4}$ for a shock radius at $r_\sh=5r_*$, to $\sim 0.06 $ for $r_\sh=2.5r_*$.

In order to be able to subtract the entropy and vorticity perturbations (described by the variables $\delta S$ and $\delta q$ defined in Appendix~\ref{sec:equations}), one needs to determine the values of the other variables $\delta f$ and $\delta h$ associated to this advected wave. This can be done as in the WKB analysis of \citet{foglizzo07,yamasaki08}, by requiring that these perturbations are advected:
\begin{eqnarray}
\frac{\p}{\p r}\left( \delta f \right)  &=& \frac{i\omega}{v_r}\delta f,    
	\label{eq:adv_df1}  \\
\frac{\p}{\p r}\left( \delta h \right)  &=&\frac{i\omega}{v_r}\delta h.  
	  \label{eq:adv_dh1}
\end{eqnarray}
Using the differential system recalled in Appendix~A to remove the radial derivatives (neglecting the non adiabatic terms in Equations~(\ref{eq:diff_dS})-(\ref{eq:diff_dq}) due the quasi-adiabatic assumption), Equations~(\ref{eq:adv_df1}) and (\ref{eq:adv_dh1}) can be expressed by the following linear system of two equations:
\begin{eqnarray}
\frac{\delta f_{\rm adv}}{c^2} - \M^2 \delta h_{\rm adv} &=& \left\lbrack1+(\gamma-1)\M^2\right\rbrack \frac{\delta S}{\gamma} - \frac{\delta q}{c^2}, 
	\label{eq:adv_df2}  \\
\mu^2\frac{\delta f_{\rm adv}}{c^2} - \delta h_{\rm adv} &=& \delta S - \frac{\delta q}{c^2}.    
	\label{eq:adv_dh2}
\end{eqnarray}

Note that Equation~(\ref{eq:adv_df2}) (signifying that $\delta f$ is advected) is equivalent to requiring that the pressure perturbation vanishes (compare Equations~(\ref{eq:adv_df2}) and (\ref{eq:dp})). Thus, when the advected wave is subtracted from the perturbations below the shock, the pressure perturbation is unchanged. This system of equations can be solved to obtain the following expression of the advected perturbations as a function of the amplitude of the entropy and vorticity perturbations (described by the variables $\delta S$ and $\delta q$):
\begin{eqnarray}
\delta f_{\rm adv} &=&  \frac{1-\M^2}{1-\mu^2\M^2}\left\lbrack \frac{c^2}{\gamma}\delta S - \delta q  \right\rbrack,   
	\label{eq:fadv}  \\
\delta h_{\rm adv} &=& \frac{1}{1-\mu^2\M^2}\bigg\lbrack (\frac{\mu^2}{\gamma}(1-\M^2) - 1)\delta S  \nonumber \\
&&+  (1-\mu^2)\frac{\delta q}{c^2}    \bigg\rbrack.
	\label{eq:hadv} 
\end{eqnarray}

In order to compute purely acoustic modes, the advected wave is then subtracted from the perturbations at the shock to define new boundary values $\delta f_{\rm acsh}$ and $\delta h_{\rm acsh}$:
\begin{eqnarray}
\delta f_{\rm acsh} &=& \delta f_\sh  - \delta f_{\rm adv},      \\
\delta h_{\rm acsh} &=&  \delta h_\sh  - \delta h_{\rm adv},
\end{eqnarray}
where $\delta f_\sh$ and $ \delta h_\sh$ are determined by the boundary conditions at the shock (Equations~(\ref{eq:dfsh}) and (\ref{eq:dfsh})). 
$\delta f_{\rm adv}$ and $\delta h_{\rm adv}$ are computed using Equations~(\ref{eq:fadv}) and (\ref{eq:hadv}), and the values of $\delta S$ and $\delta q$ at the shock (Equations~(\ref{eq:dSsh}) and (\ref{eq:dqsh})). Using these modified boundary conditions, the acoustic modes can then be computed numerically as described for example in \citet{foglizzo07}. 

Figure~\ref{fig:acoustic_growth_rate} compares the growth rate of the most unstable acoustic modes (left panel) and SASI modes (right panel). The difference is striking: the purely acoustic modes are all stable for all shock radii in the explored range $r_\sh=2-100r_*$, while SASI modes have significant growth rates. This shows that the purely acoustic cycle cannot account for SASI and that the advected wave is essential to the instability. The most "unstable" acoustic mode (actually the least damped) for each $l$ is the lowest frequency one. The discussion of the purely acoustic frequency spectrum is delayed to Section~\ref{sec:SASI_spectrum}.

In order to quantify the effect of non-adiabaticity at the shock on the acoustic modes, we have used boundary conditions which either take into account or neglect the cooling processes in the computation of $\delta S$ and $\delta q$. The associated change in the frequency and growth rate of the modes is barely noticeable (less than $2\%$ for any shock radius), confirming the validity of our quasi-adiabatic approximation.

The error due to the high frequency approximation can in turn be quantified in the following way. Noting that there is some arbitrariness in the choice of the variables that should be advected, one can require another variable to be advected and compare the result with the previous calculation. For example requiring the perturbation of radial velocity $\delta v_r/v_r$ to be advected instead of $\delta h$, we get a different definition of the advected perturbations:
\begin{eqnarray}
\frac{\delta f_{\rm adv}}{c^2} &=& \frac{1}{1+\epsilon_1 -\M^2(\mu^2+\epsilon_2)} \bigg\lbrack (1+\epsilon_1)(1-\M^2)\frac{\delta S}{\gamma}  \nonumber   \\
&& -\left(1+\epsilon_1 - \M^2(1+\epsilon_2)\right)\frac{\delta q}{c^2}   \bigg\rbrack  	
	\label{eq:fadv_bis}   \\
\delta h_{\rm adv} &=& \frac{\mu^2+\epsilon_2}{1+\epsilon_1}\frac{\delta f_{\rm adv}}{c^2} - \delta S - \frac{1+\epsilon_2}{1+\epsilon_1}\frac{\delta q}{c^2} 	
	\label{eq:hadv_bis} 
\end{eqnarray}
where $\epsilon_1$ and $\epsilon_2$ are two parameters vanishing in the limit  $\omega r_\sh/v_\sh \gg 1 $, and defined by:
\begin{eqnarray}
\epsilon_1 &\equiv& -\frac{v}{i\omega}\p_r \M^2  \\
\epsilon_2 &\equiv& \frac{v}{i\omega}\frac{1}{c^2}\p_r(c^2-v^2)
\end{eqnarray}
As expected, the two prescriptions for the advected perturbations are equivalent in the limit of vanishing $\epsilon_1$ and $\epsilon_2$. These parameters are very small for the second and third $l=1$ acoustic modes: $\epsilon_1 \sim \epsilon_2 \sim 0.03$ and $0.015$ respectively for $r_\sh=2.5r_*$ (note that these two modes have a higher frequency than the second and third SASI modes as discussed in Section~\ref{sec:SASI_spectrum}). The high frequency approximation is thus well justified and the two prescriptions are in reasonable agreement: the damping rates agree within $10-15\%$ for the second mode, and $<5\%$ for the third mode. These two modes are enough to conclude that SASI is not a purely acoustic instability since the second and third $l=1$ SASI modes are unstable. The approximation is slightly less good for the fundamental acoustic mode: in this case $\epsilon_1 \sim \epsilon_2 \sim 0.1$. The damping rates in the two prescriptions differ by $\sim 30\%$, but the modes are still definitely stable whatever the prescription.

\section{Detailed analysis of the advective-acoustic and purely acoustic frequency spectra}
	\label{sec:details}

In this section, we study in detail the frequency spectrum expected from advective-acoustic and purely acoustic cycles. In the subsection~\ref{sec:toy_model}, we study a simple toy model \citep{foglizzo09}, which allows an analytical study of the frequency spectrum. In the following subsection~\ref{sec:spherical}, we then give an approximate generalisation to the more realistic model of SASI used in Sections~\ref{sec:radial_time} and \ref{sec:acoustic_modes}. This analysis confirms the validity of the argument presented in Section~\ref{sec:radial_time}. Furthermore it allows a direct comparison between  analytical expectations and the frequency spectrum of SASI. This direct comparison leads to the same conclusion about the instability mechanism, in favour of the advective-acoustic cycle.

\subsection{Planar toy model}
	\label{sec:toy_model}
\subsubsection{Presentation}

We consider the toy model introduced by \citet{foglizzo09}. It consists of a planar shock lying above a zone of deceleration imposed by an external potential step. The potential step is localised within a length scale of $H_\nabla$, and is separated from the shock by a region of size $H$ where the flow is uniform. The gas is assumed to be perfect, adiabatic (with an adiabatic index $\gamma=4/3$), and non-self-gravitating. The direction perpendicular to the shock is called $z$, and referred to as vertical. Modes have a transverse structure along $x$ (parallel to the shock), which is referred to as the horizontal direction. The domain is a box of horizontal size $L$. After a Fourier transform in time and horizontal direction $x$, the eigenmodes are characterised by their complex frequency $\omega$ and the number $n_x$ of horizontal wavelengths in the width of the box. The horizontal wave number is then:
\begin{equation}
k_x = \frac{2\pi n_x}{L}.
	\label{eq:kx}
\end{equation}
The reader is referred to \citet{foglizzo09} for more details about this toy model, the equations governing it, and the numerical method used to solve them. In the remaining of this subsection, the numerical calculations have used the parameters: $L=6H$, $\M_1=\infty$, $H_\nabla/H=0.01$, and $c_{\rm in}^2/c_{\rm out}^2 = 0.75$.

Cycle constants $Q$ and $R$ describing the advective-acoustic and purely acoustic cycle can be defined as in Section~\ref{sec:review} (but here without any approximation). An eigenmode with a complex eigenfrequency $\omega$ must then verify Equation~(\ref{eq:QR}), and contains both cycles in some proportion. In order to separate the two cycles and study their properties independently from each other, we follow \citet{foglizzo09} in considering advective-acoustic modes verifying:
 \begin{equation}
Q(\omega) = 1,
	\label{eq:Q}
\end{equation}
which corresponds to a situation where the acoustic wave propagating downward would be artificially removed. Similarly, purely acoustic modes follow:
\begin{equation}
R(\omega) = 1,
	\label{eq:R}
\end{equation}
which signifies that the advected entropy-vorticity wave is ignored. This is equivalent to the method described in Section~\ref{sec:acoustic_modes} to compute purely acoustic modes. 

As in \citet{foglizzo09}, we define new cycle constants $Q_0$ and $R_0$ that do not include the propagation of the waves between the shock and the potential step. These are defined as:
\begin{eqnarray}
Q &=& Q_0e^{i(k_z^{\rm adv}-k_z^-)H}, \label{eq:Q0}\\
R &=& R_0e^{i(k_z^{+}-k_z^-)H}, \label{eq:R0}
\end{eqnarray}
where $k_z^{\rm adv}$ and $k_z^\pm$ are the vertical numbers of the advected wave, and of the acoustic waves propagating down (+) and up (-):
\begin{eqnarray}
k_z^{\rm adv} &=& \frac{\omega}{v_z}, \\
k_z^\pm &=& \frac{\omega}{c}\frac{\M\mp\mu}{1-\M^2}, \\
\mu^2 &=& 1 - \frac{k_x^2c^2(1-\M^2)}{\omega^2}. 
\end{eqnarray}

\subsubsection{Eigenfrequencies of the advective-acoustic cycle}
Equations~(\ref{eq:Q}) and (\ref{eq:Q0}) are combined to obtain the condition verified by an advective-acoustic mode:
\begin{equation}
Q_0e^{i\omega\tau_{\rm aac}\frac{1+\mu\M}{1+\M}} = 1,
	\label{eq:Qmode}
\end{equation}
where $\tau_{\rm aac}$ is the duration of a one dimensional advective-acoustic cycle ($k_x = 0$, $\mu = 1$):
\begin{equation}
\tau_{\rm aac} \equiv \frac{H}{|v|} + \frac{H}{c(1-\M)} = \frac{H}{|v|(1-M)}.
	\label{eq:taac}
\end{equation}
Equation~\ref{eq:Qmode} can be solved approximately in the regime where the growth rate is low ($\omega_i \ll \omega_r$) and where the acoustic wave is propagating ($\mu^2>0$). In this case, the modulus of this complex relation gives access to the growth rate \citep{foglizzo09}:
\begin{equation}
\omega_i \simeq \frac{1+\M}{\mu_r+\M}\mu_r \frac{\log |Q|}{\tau_{\rm aac}}.
	\label{eq:wi_aac}
\end{equation}
In a similar way, the frequency may be obtained from the argument of the same relation, which gives an information on the wave phase:
\begin{equation}
\Phi(Q_0) + Re\left(\omega\tau_{\rm aac}\frac{1+\mu\M}{1+\M}\right) = 2n_z\pi,
	\label{eq:phaseQ}
\end{equation}
where $\Phi(Q_0)$ is the phase of $Q_0$, and describes the phase shift taking place during the two coupling processes. Equation~(\ref{eq:phaseQ}) gives a second order polynomial in $\omega_r$, which physical solution is:
\begin{eqnarray}
\omega_r &=& \frac{1}{1-\M} \bigg\lbrack (n_z-\Phi/2\pi)\omega_{\rm aac}   \nonumber \\
 && -\M\left\lbrack(n_z-\Phi/2\pi)^2\omega_{\rm aac}^2 - n_x^2\omega_{{\rm ac}x}^2(1-\M)^2\right\rbrack^{1/2} \bigg\rbrack,
	\label{eq:w_aac}
\end{eqnarray}
where we defined two frequencies typical of the advective-acoustic cycle and of the horizontal propagation:
\begin{eqnarray}
\omega_{\rm aac} &\equiv& \frac{2\pi}{\tau_{\rm aac}}, \\
\omega_{{\rm ac}x} &\equiv& k_xc.
\end{eqnarray}
Note that in principle $\Phi$ depends on the frequency, therefore Equation~(\ref{eq:w_aac}) gives explicitly the frequency only if we know the value of $\Phi$. In the following of this subsection, we suppose $\Phi=0$ which turns out to be a good approximation in this toy model. This is not always true in the spherical model as will be discussed in Section~\ref{sec:SASI_spectrum}.

\begin{figure*}
\includegraphics[width=\columnwidth]{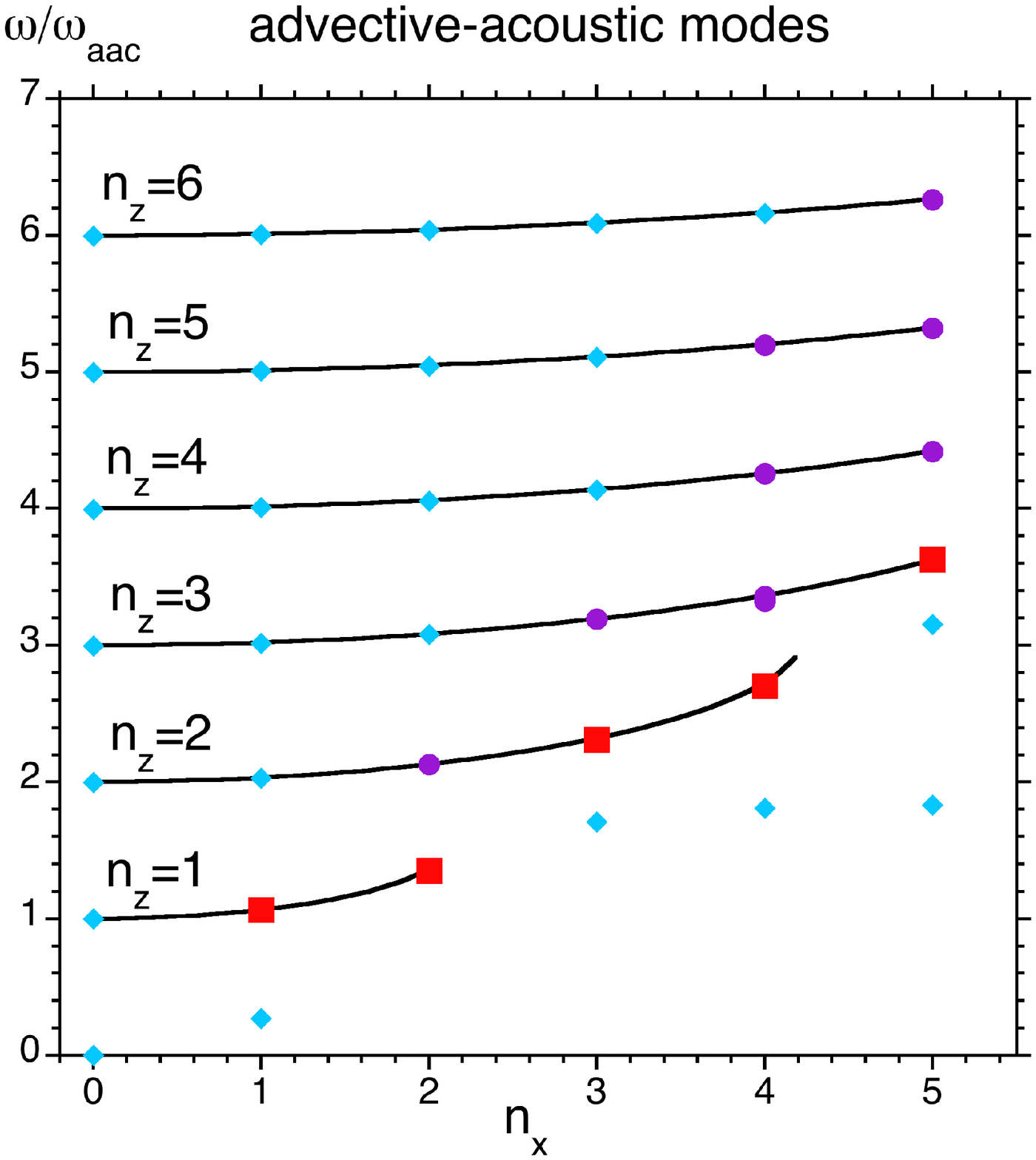}
\includegraphics[width=\columnwidth]{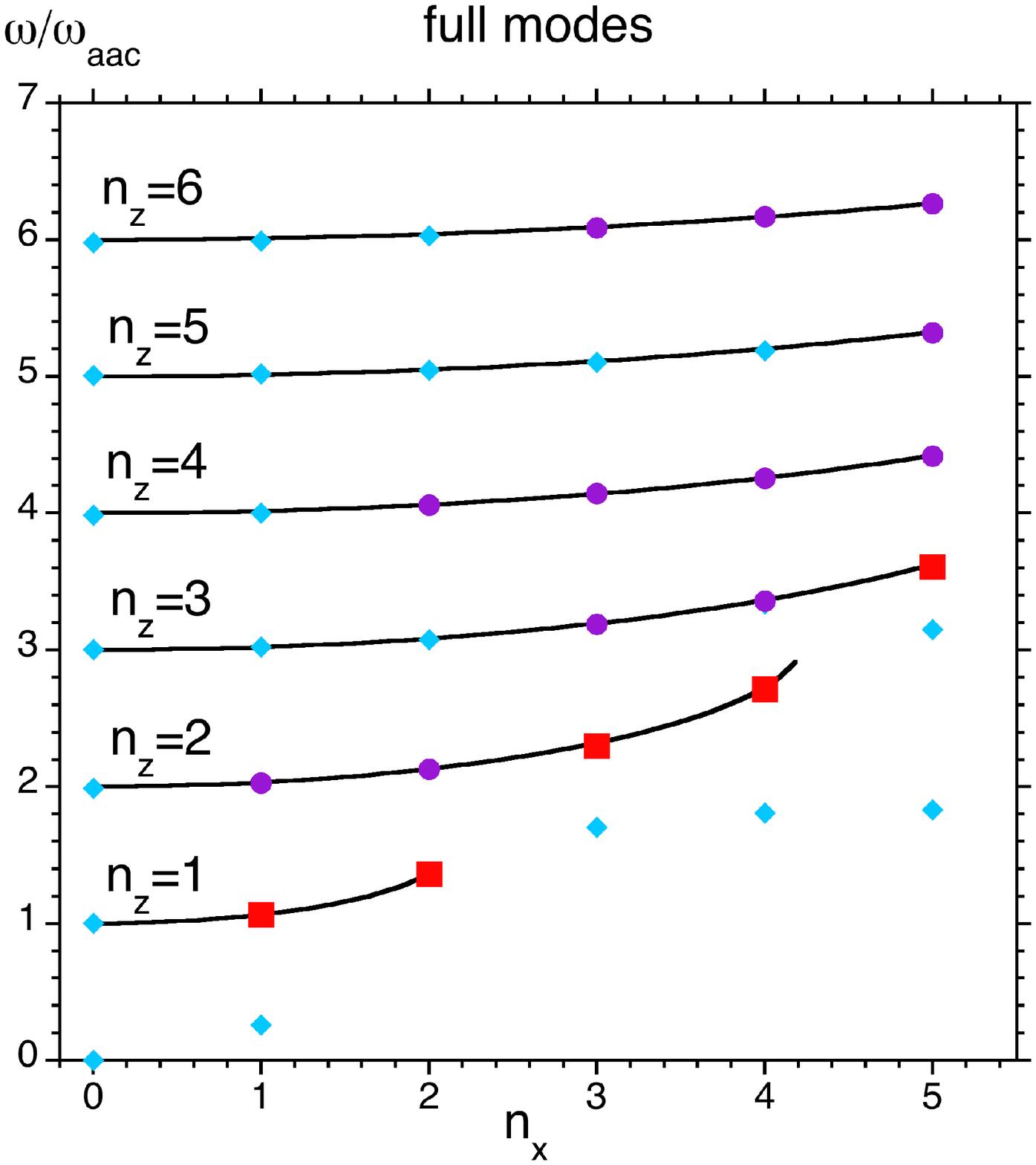}
	\caption{Eigenspectrum of the advective-acoustic cycle in the planar toy model of \citet{foglizzo09}: oscillation frequency as a function of $n_x$ (the number of transverse wavelength in the width of the box). Symbols represent the numerically computed eigenmodes for the advective-acoustic cycle only (left panel) and the full toy model (right panel). The colour and shape of the symbols reveal minor differences in their stability properties: turquoise diamonds for stable, purple circles for unstable, and red squares for the most unstable mode for a given value of $n_x$. Continuous lines represent the analytical solution given by Equation~(\ref{eq:w_aac}) and assuming $\Phi=0$ (the curves are restricted to the regime of propagating acoustic waves, where the analytical expression applies). The analytical expression is in very good agreement with the numerical modes in both panel, as would be expected from the fact that the advective-acoustic cycle is driving the instability in this toy model. }
		\label{fig:toyspectrum_aac}
\end{figure*}

\begin{figure*}
\centering
\includegraphics[width=\columnwidth]{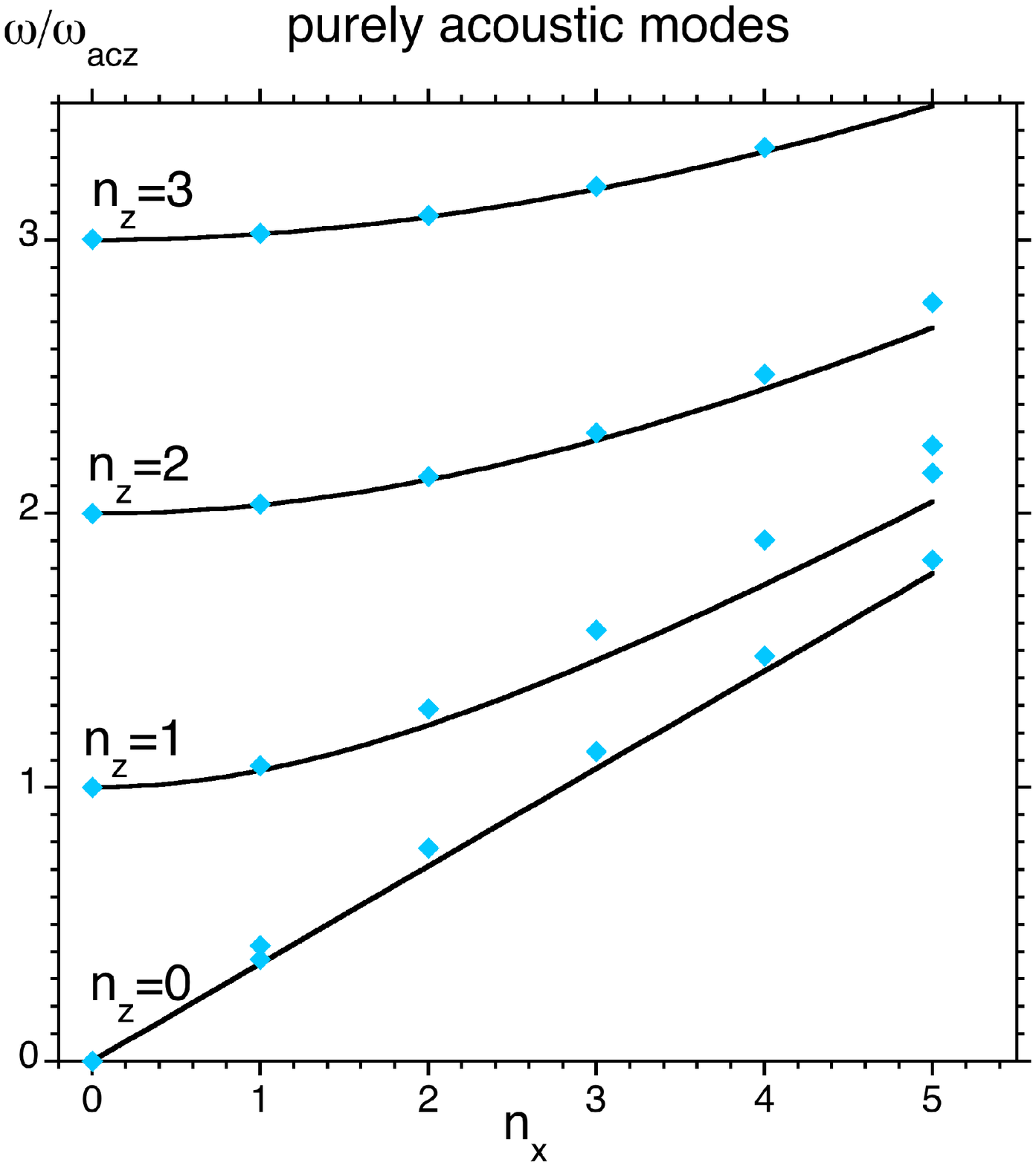}
\includegraphics[width=\columnwidth]{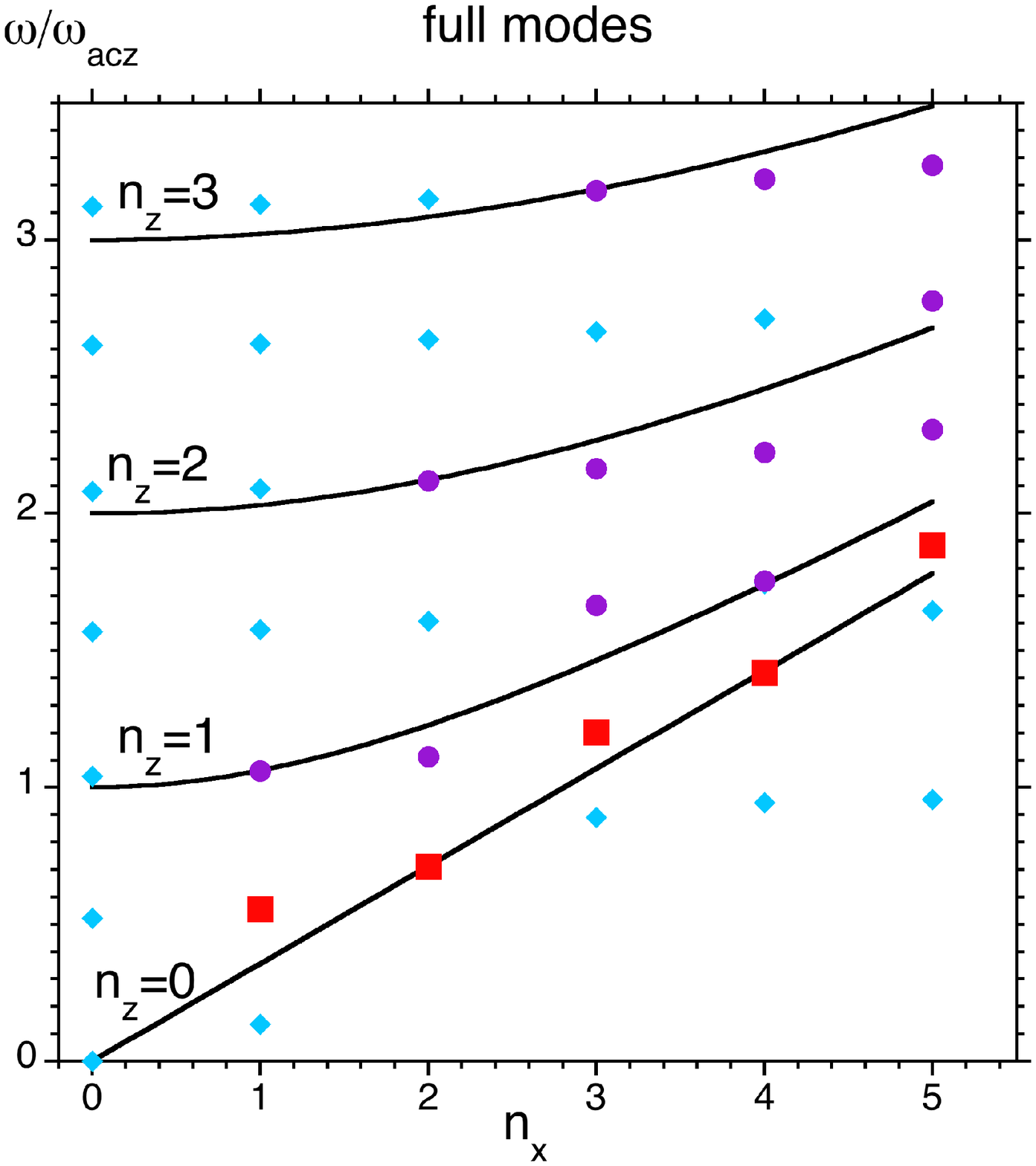}
	\caption{Same as Figure~\ref{fig:toyspectrum_aac} but for the purely acoustic cycle. The analytical expression (Equation~(\ref{eq:w_acoustic}), with $\Phi=0$) agrees with the purely acoustic modes (left panel), but not with the full modes (right panel) confirming that these are not dominated by the purely acoustic cycle.}
	\label{fig:toyspectrum_acoustic}
\end{figure*}

Figure~\ref{fig:toyspectrum_aac} compares this analytical result with eigenmodes computed numerically. The agreement is very good both with advective-acoustic modes and with the full modes (i.e. the modes that verify Equation~(\ref{eq:QR})), in the regime of propagative acoustic waves. The left panel shows that the approximations made are valid, and that there is no significant phase shift. The right panel shows that the frequency of the full modes is dictated by the advective-acoustic cycle, as could be expected from the results of \citet{foglizzo09}. One can nonetheless notice that the purely acoustic cycle has an influence on the growth rate of the modes (especially those close to marginal stability, \citep{foglizzo09}), since the unstable modes are not exactly the same on the two panels. The origin of the stable $n_x=1$ mode at very low frequency, which does not coincide with any curve, is rather uncertain (it lies in the regime of evanescent acoustic waves, which is not described by Equation~(\ref{eq:w_aac})).

From Figure~\ref{fig:toyspectrum_aac} and Equation~(\ref{eq:w_aac}), one may notice that two successive modes sharing the same transverse structure are separated by a frequency difference that is close to the frequency of a vertical advective-acoustic cycle $\omega_{\rm aac}$. This is exact in the case of one-dimensional modes, but only approximative for $n_x\neq 0$. Indeed a transverse advective-acoustic mode has a slightly higher frequency than the corresponding 1D mode (by a factor at most $1/(1-\M)$ for a propagative mode). This higher frequency may seem counterintuitive at first sight, because the geometrical path along which the waves propagate during one cycle is longer. This longer path is the reason why the duration of the cycle increases with inclination in Equation~(\ref{eq:wi_aac}) giving the growth rate. However this duration corresponds to the group velocity of acoustic waves, while the frequency is determined by a phase relation. The cycle time relevant to the frequency is thus given by the phase speed. The vertical phase speed of an acoustic wave increases with inclination, because the wave fronts get closer to the vertical direction. This explains the higher frequency of non radial modes.

\subsubsection{Eigenfrequencies of the purely acoustic cycle}
The frequency of the purely acoustic modes can be obtained by a very similar method to that used for the advective-acoustic modes. For this purpose, one combines Equations~(\ref{eq:R}) and (\ref{eq:R0}) to obtain:
\begin{equation}
R_0e^{i\omega\tau_{{\rm ac}z}\mu} = 1,
	\label{eq:Rmode}
\end{equation}
where $\tau_{{\rm ac}z}$ is the duration of a one-dimensional purely acoustic cycle ($k_x=0$):
\begin{equation}
\tau_{{\rm ac}z} \equiv \frac{H}{c(1+\M)} + \frac{H}{c(1-\M)}= \frac{2H}{c(1-\M^2)}.
\end{equation}
By taking the argument of this complex equation, we obtain:
\begin{equation}
\Phi(R_0) + Re(\tau_{{\rm ac}z}\sqrt{\omega^2 - n_x^2\omega_{{\rm ac}x}^2(1-\M^2)}) = 2n_z\pi.
\end{equation}
from which one can deduce the following expression for the frequency of purely acoustic modes:
\begin{equation}
\omega_r = \left\lbrack (n_z-\Phi/2\pi)^2\omega_{{\rm ac}z}^2 + n_x^2\omega_{{\rm ac}x}^2(1-\M^2) \right\rbrack^{1/2},
	\label{eq:w_acoustic}
\end{equation}
where $\omega_{{\rm ac}z}$ is the frequency associated to a vertical acoustic propagation:
\begin{equation}
\omega_{{\rm ac}z} \equiv \frac{2\pi}{\tau_{{\rm ac}z}}.
	\label{eq:wacz}
\end{equation}
Equation~(\ref{eq:w_acoustic}) is analogous to the frequency of the acoustic modes of a rectangular box with periodic boundary conditions in $x$ and reflexive boundary conditions in $z$  (it is only slightly modified by advection):
\begin{equation}
\omega = 2\pi c^2\left\lbrack\frac{n_z^2}{(2H)^2} + \frac{n_x^2}{L_x^2}\right\rbrack.
	\label{eq:w_box}
\end{equation}
Thus as argued in Section~\ref{sec:radial_time}, if the number of wavelengths in the transverse direction ($n_x$) increases, the frequency increases typically by $\omega_{{\rm ac}x}$ corresponding to a propagation time in the transverse direction. Similarly if the number of wavelength in the vertical direction ($n_z$) increases, the frequency increases typically by $\omega_{{\rm ac}z}$ corresponding to an acoustic propagation along the vertical direction.

Equation~(\ref{eq:w_acoustic}) (with $\Phi=0$) is compared with the numerically computed eigenmodes in Figure~\ref{fig:toyspectrum_acoustic}. It agrees with the purely acoustic modes (all stable), showing here again that the approximation of small growth rate holds and that the phase shift is not very significant. By contrast, the frequency of the full modes do not match the purely acoustic prediction. This confirms that the unstable modes are not dominated by the purely acoustic cycle, and that the study of the frequency spectrum enables to discriminate between the two mechanisms. 

Although the purely acoustic cycle does not determine the mode frequencies, it can be observed from Figure~\ref{fig:toyspectrum_acoustic} that the acoustic timescale influences the modes stability. Indeed, the full modes in the vicinity of an acoustic branch are more likely to be unstable than the other modes. Furthermore, the most unstable modes (for a given $n_x$) are all close to the branch $n_z=0$. These observations can be interpreted by two phenomena. First, a constructive interference between the purely acoustic and advective-acoustic cycles takes place when the mode frequency (determined by the advective-acoustic cycle) is close to that of a purely acoustic cycle \citep{foglizzo02}. This destabilising effect is more pronounced for modes close to marginal stability \citep{foglizzo09}. Second, the growth rate associated with the advective-acoustic cycle alone is maximum when the acoustic propagation is close to horizontality \citep{foglizzo09}, which happens at the frequency of the fundamental acoustic branch $n_z=0$.

\subsection{Spherical model}
\subsubsection{Approximate generalisation of the analytical results}
	\label{sec:spherical}

The description of waves propagation given in the previous subsection is exact in a uniform flow, and could be generalised to a non uniform flow in spherical geometry with the use of the WKB approximation. In this framework, the terms of the form $e^{ik_zH}$ (in Equations~(\ref{eq:Q0}) and \ref{eq:R0}) should be replaced by terms of the following form: $e^{i\int k_r\, \dd r}$. However, as our goal is to discriminate the mechanism at work in the regime where the WKB approximation does not hold, we will not develop rigorously such a description. We simply give an approximate generalisation, from which one may expect the right order of magnitude but not an exact value. This is however sufficient to distinguish the two mechanisms as noted in Section~\ref{sec:radial_time}.

The timescales associated to a radial advective-acoustic cycle, a radial purely acoustic cycle, and a transverse acoustic propagation have been defined for a spherical model in Section~\ref{sec:timescales} (Equations~(\ref{eq:taac_spherical})-(\ref{eq:tacphi})). One can define from these timescales, typical frequencies which are a direct generalisation of the frequencies of the planar model $\omega_{\rm aac}$, $\omega_{{\rm ac}z}$, and $\omega_{{\rm ac}x}$:
\begin{eqnarray}
\omega_{\rm aac} &\equiv& \frac{2\pi}{\tau_{\rm aac}}, \\
\omega_{{\rm ac}r}&\equiv& \frac{2\pi}{\tau_{{\rm ac}r}}, \\
\omega_{{\rm ac}\phi}(r) &\equiv&  \frac{2\pi}{\tau_{{\rm ac}\phi}} = \frac{c}{r}.
\end{eqnarray}
We propose to estimate the frequency spectra of advective-acoustic and purely acoustic cycles by directly introducing these typical frequencies in Equations~(\ref{eq:w_aac}) and (\ref{eq:w_acoustic}) (for this purpose $n_x^2\omega_x^2$ should be replaced by $l(l+1)\omega_{{\rm ac}\phi}^2$). This procedure suffers from different sources of uncertainty: the formula itself is approximate, the phase shift is {\it a priori} unknown, and the typical frequencies $\omega_{\rm aac}$, $\omega_{{\rm ac}r}$, and $\omega_{{\rm ac}\phi}$ are uncertain respectively due the lack of knowledge of the coupling radius, the turning point of acoustic waves, and the radius at which it should be evaluated. When comparing to SASI modes in Section~\ref{sec:SASI_spectrum}, we will use a range of values for the radius where $\omega_{{\rm ac}\phi}$ and $\M$ are evaluated in order to illustrate this uncertainty.

The above formulation can be used to assess the validity of the argument developed in Section~\ref{sec:radial_time}. An inspection of Figures~\ref{fig:toyspectrum_aac}, \ref{fig:toyspectrum_acoustic}, \ref{fig:acoustic_spectrum} and \ref{fig:SASI_spectrum} confirms visually that the difference between two successive modes that share the same transverse structure corresponds approximately to a radial propagation frequency: $\omega_{\rm aac}$ for the advective-acoustic cycle, and $\omega_{{\rm ac}r}$ for the purely acoustic cycle. This visual observation can be recovered from Equation~(\ref{eq:w_aac}) for the advective-acoustic cycle, if one assumes $l(l+1)\omega_{{\rm ac}\phi}^2 \ll n_r^2\omega_{\rm aac}^2 $:
\begin{eqnarray}
\Delta \omega & \simeq  & \left(1 - \frac{\Phi^{\prime}- \Phi}{2\pi} \right)\omega_{\rm aac} \bigg\lbrack 1 \nonumber \\
&&- \frac{\M(1-\M)l(l+1)\omega_{{\rm ac}\phi}^2}{2(n_r -\Phi/2\pi)(n_r+1 -\Phi^{\prime}/2\pi)\omega_{\rm aac}^2} + ...   \bigg\rbrack
	\label{eq:domega_aac} 
\end{eqnarray}
where we have defined $\Delta \omega \equiv \omega_r(n_r+1) - \omega_r(n_r)$, $\Phi \equiv \Phi(n_r)$, and $\Phi^{\prime}\equiv \Phi(n_r+1)$. Similarly for the purely acoustic cycle, Equation~(\ref{eq:w_acoustic}) with the assumption that $l(l+1)\omega_{{\rm ac}\phi}^2 \ll n_r^2\omega_{{\rm ac}r}^2 $ leads to the following relation:
\begin{eqnarray}
\Delta \omega & \simeq  & \left(1 - \frac{\Phi^{\prime}- \Phi}{2\pi} \right) \omega_{{\rm ac}r}  \bigg\lbrack 1 \nonumber \\
&& - \frac{(1-\M^2)l(l+1)\omega_{{\rm ac}\phi}^2}{2(n_r -\Phi/2\pi)(n_r+1 -\Phi^{\prime}/2\pi)\omega_{{\rm ac}r}^2} + ...   \bigg\rbrack 
	\label{eq:domega_acoustic}
\end{eqnarray}
Thus we expect the timescale extracted from the eigenfrequencies in Section~\ref{sec:radial_time} to match one of the radial times (depending on the instability mechanism), for sufficiently large values of $n_r$ and if the two modes have the same phase shift. The deviation at low $n_r$ can be estimated from the second term in the right hand side of Equations~(\ref{eq:domega_aac}) and (\ref{eq:domega_acoustic}). For example for the parameters $r_\sh = 2.5r_*$, $l=1$, $n_r=1$, it is already less than $\sim 10\%$ for both the advective-acoustic and the purely acoustic modes (note that for the purely acoustic modes $n_r=1$ is the second mode). This confirms the validity of the argument given in Section~\ref{sec:radial_time}, and is roughly consistent with the fact that the three extracted times differ by at most $15-20\%$. We note that a difference in the phase shift between two successive modes could pollute the measure of the radial time, but it is not expected to consistently change this measure for different mode duplets.

\subsubsection{Comparison with SASI and purely acoustic eigenfrequencies}
	\label{sec:SASI_spectrum}

\begin{figure}
\centering
\includegraphics[width=\columnwidth]{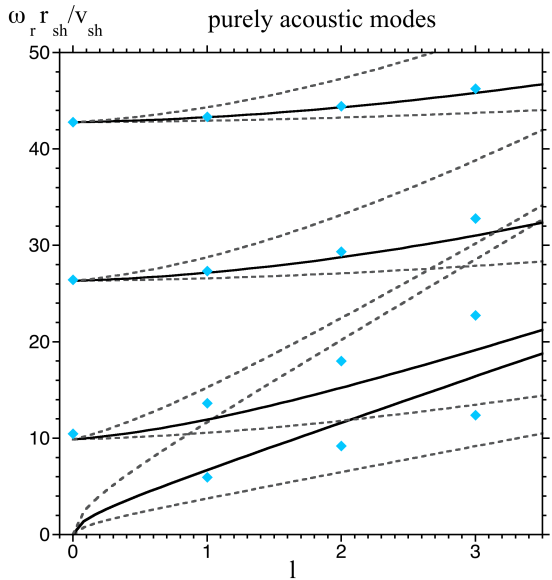}
	\caption{Comparison of the frequency spectrum of the purely acoustic eigenmodes (computed as described in Section~\ref{sec:acoustic_modes}) with the analytical estimate, in the spherical model of \citet{blondin06}. The frequency in units of $v_\sh/r_\sh$ is represented as a function of the spherical harmonics index $l$. Symbols represent eigenmodes with the same colour and shape convention as the previous figures. Full lines correspond to analytical estimates in which $\M$ and $\omega_{{\rm ac}\phi}$ are calculated at the radius $r=(r_\sh-r_*)/2$, while dashed lines use $r=r_\sh$ and $r=1.1r_*$. In order to obtain a decent agreement, the phase shift of the harmonics ($n_r>0$) had to be adjusted to the value $\Phi/(2\pi) = 0.4$. }
	\label{fig:acoustic_spectrum}
\end{figure}

\begin{figure*}
\centering
\includegraphics[width=\columnwidth]{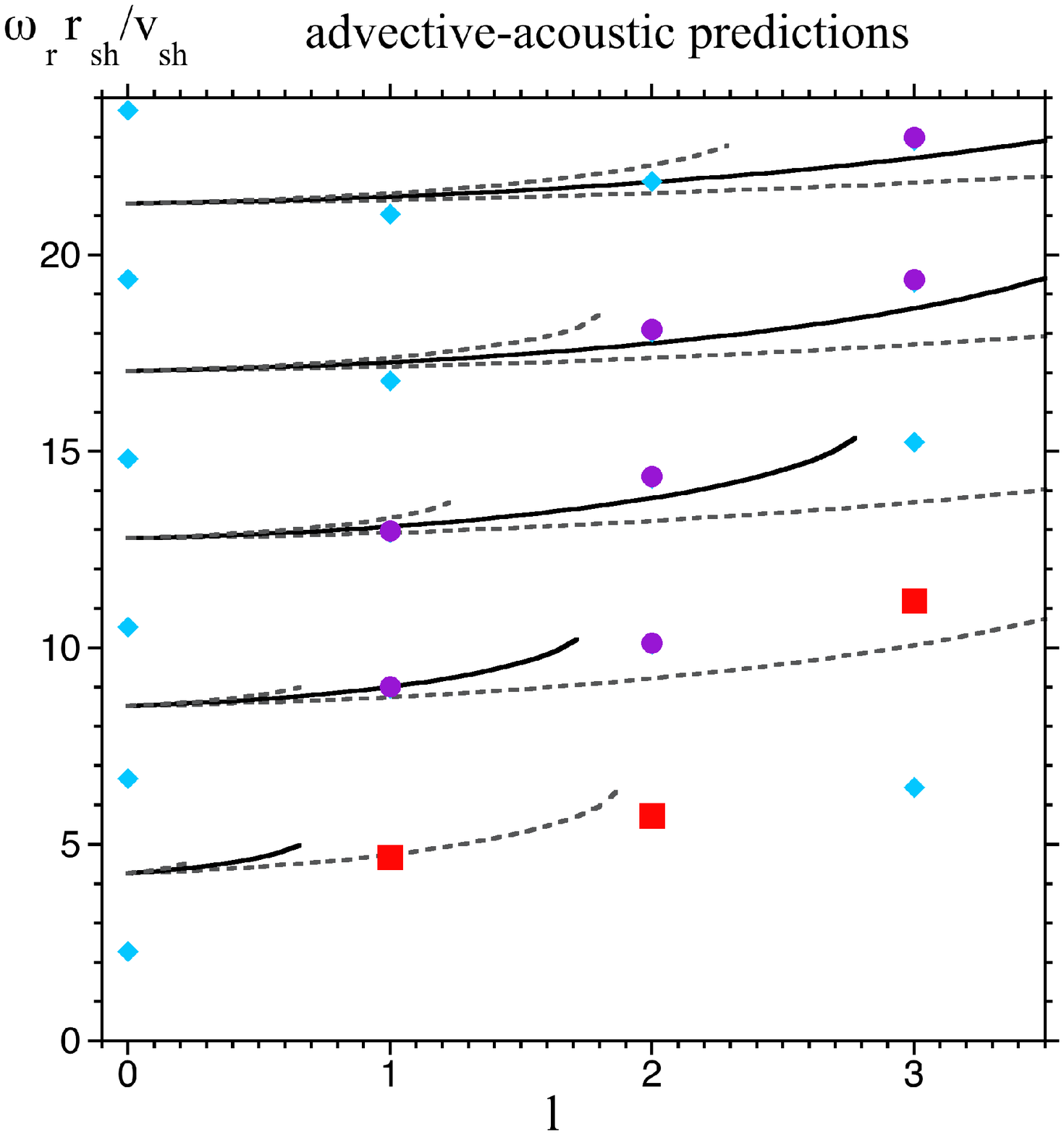}
\includegraphics[width=\columnwidth]{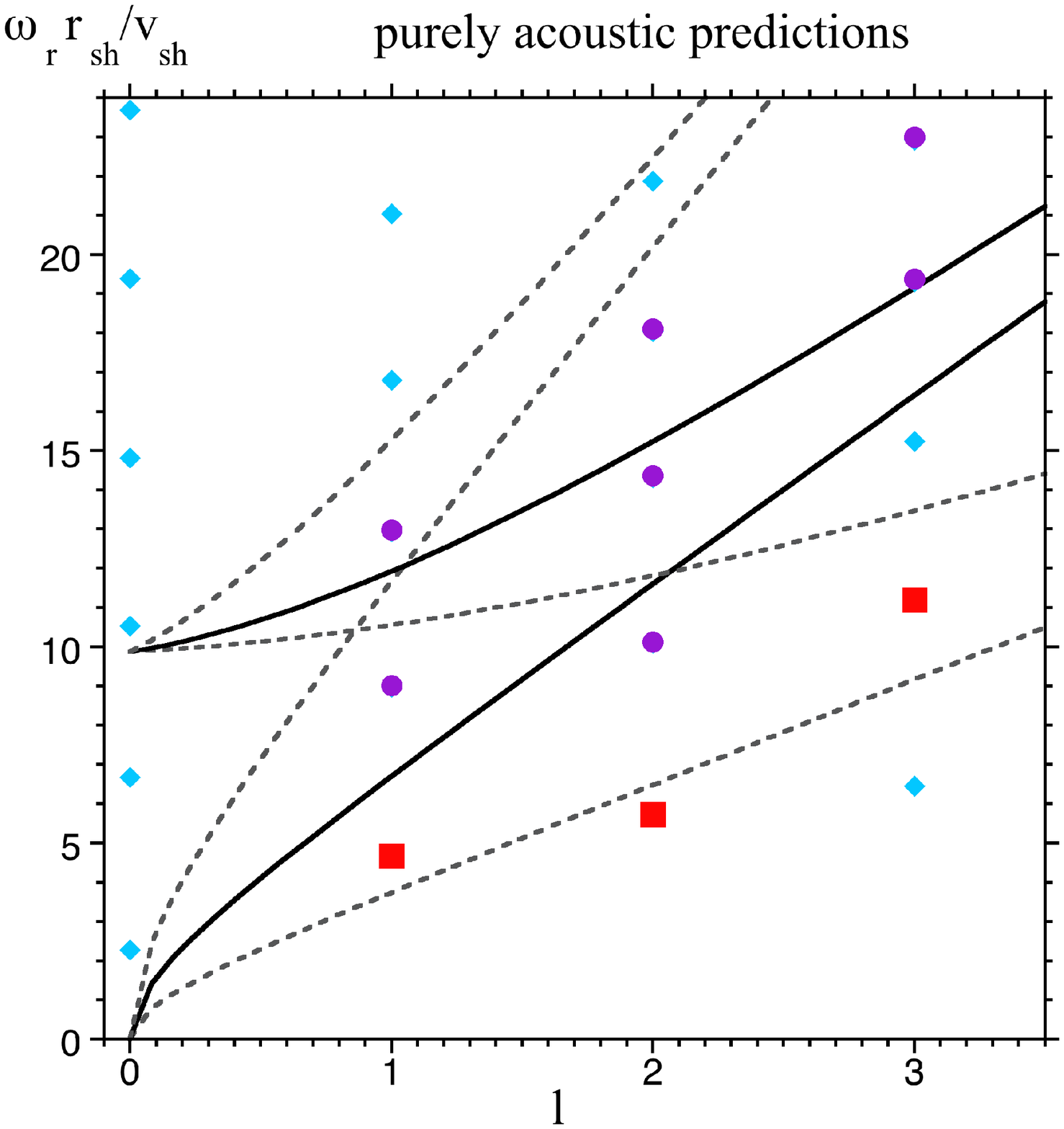}
	\caption{Comparison of the frequency of SASI eigenmodes with the estimates for an advective-acoustic cycle (left panel) and a purely acoustic cycle (right panel). The frequency in units of $v_\sh/r_\sh$ is represented as a function of the spherical harmonics index $l$. Symbols represent eigenmodes with the same colour and shape convention as the previous figures. Full lines correspond to analytical estimates {\bf in which $\M$ and $\omega_{{\rm ac}\phi}$ are calculated at the radius $r=(r_\sh-r_*)/2$,} while dashed lines use $r=r_\sh$ and $r=1.1r_*$. As previously, the advective-acoustic frequency estimate is represented only in the regime of propagating acoustic waves (at the radius $r$ used for the estimate). The prediction from the advective-acoustic cycle is roughly in agreement with all the non radial modes. The purely acoustic cycle approximately predicts the frequency of the most unstable modes (for a given $l$), but strongly overestimates the frequency of higher frequency harmonics. }
	\label{fig:SASI_spectrum}
\end{figure*}

Figure~\ref{fig:acoustic_spectrum} compares the frequency spectrum of purely acoustic modes with the analytical estimate. The method described in Section~\ref{sec:acoustic_modes} gives two kinds of modes: the acoustic modes in which we are interested, as well as modes due to a secondary advective-acoustic cycle with the entropy  being created by cooling processes acting on the acoustic wave. This last category of modes are clearly distinguished and significantly more stable than the purely acoustic modes, and therefore we do not show them in this discussion of the purely acoustic modes. Contrary to the case of the planar toy model, it is necessary to invoke a phase shift so that the analytical estimate agrees with the numerically obtained eigenfrequencies. We find that a phase shift of $\Phi/(2\pi)=0.4$ for all modes with $n_r>0$, and $\Phi=0$ for all modes with $n_r = 0$ gives a reasonable match to the eigenspectrum (Figure~\ref{fig:acoustic_spectrum}). The origin of this phase shift remains obscure. Nevertheless, the comparison confirms that the analytical estimate accounts for the frequency of acoustic modes within the uncertainties discussed in the previous subsection and illustrated by the dashed lines in Figure~\ref{fig:acoustic_spectrum}.

Figure~\ref{fig:SASI_spectrum} compares the frequency of SASI eigenmodes with the predictions from the advective-acoustic cycle (left panel) and the purely acoustic cycle (right panel). The advective-acoustic cycle prediction agrees with the frequency of all the non radial modes, with a precision of ten to a few tens of percents. On the other hand, the purely acoustic cycle approximately predicts the frequency of the most unstable modes for each $l$, but significantly overestimates that of other modes. This is consistent with the result of Section~\ref{sec:radial_time} that the acoustic radial time is significantly shorter than that extracted from the eigenspectrum. It is also consistent with the fact that the SASI eigenspectrum is much more densely populated than the acoustic eigenspectrum (compare Figure~\ref{fig:acoustic_spectrum} and Figure~\ref{fig:SASI_spectrum}, remembering that the frequency scale is different).

Similarly to the toy model, the most unstable SASI modes for each spherical harmonics index $l$ are close to the purely acoustic branch $n_r=0$. This can again be interpreted by a higher efficiency of the advective-acoustic cycle, as well as a constructive interference with the purely acoustic mode. It is also probably linked with the observation of \citet{fernandez09a} that the growth rate of a mode is maximum when the advection time coincides with the azimuthal acoustic time.

Finally, we observe that the (stable) radial modes are systematically shifted with respect the theoretical curve. This can be explained by a phase shift taking place during the advective-acoustic cycle, as shown by \citet{fernandez09b}. Indeed, a phase shift of $\Phi=\pi$ then gives the following frequency spectrum for the $l=0$ modes:
\begin{equation}
\omega_r = (n_r - 1/2)\omega_{\rm aac},
\end{equation}
which is in much better agreement with the numerical eigenmodes.

\section{Conclusion} 
	\label{sec:conclusion}
Two new arguments have been presented which allow us to conclude that SASI cannot be explained by a purely acoustic mechanism, and should thus be interpreted as an advective-acoustic cycle. Contrary to previous works, these two arguments do not rely on a WKB analysis of the acoustic waves, and are therefore valid for realistic values of the shock radius.

First we proposed a method to extract a radial propagation timescale from the frequency difference between two successive SASI modes. This timescale agrees fairly well with the advective-acoustic time, but strongly disagrees with the radial acoustic time. This discards the purely acoustic cycle as the driving mechanism of SASI, and argues in favour of the advective-acoustic cycle. A detailed analysis of the frequency spectrum provided in Section~\ref{sec:details} confirmed the validity of this argument. 

Second, we described a new method to compute purely acoustic modes by artificially subtracting the advected wave created by the shock. All these acoustic modes were found to be stable, thus showing that the advected entropy-vorticity wave created by the oscillation of the shock plays a central role in the instability mechanism.

As a side product, we provided an analytical description of the frequency spectra of the advective-acoustic and purely acoustic cycles. This description was checked in the planar toy model of \citet{foglizzo09} and the model of \citet{blondin06}. The similarity between the frequency spectra in these two models, supports the relevance of the toy model for core collapse supernovae. Finally, we note that unexplained phase shifts during the coupling processes were observed in the second model. This calls for a better understanding of the coupling process in a cooling layer.

\section*{Acknowledgments}

The authors are grateful to John Blondin for stimulating discussions. This work has been partially funded by the Vortexplosion project ANR-06-JCJC-0119 and the SN2NS project ANR-10-BLAN-0503. J.G. acknowledges support by STFC.

\appendix

\section{Differential system and boundary conditions with the new variables of Yamasaki \& Foglizzo}
	\label{sec:equations}
Following \citet{yamasaki08}, we adopt the variables $\delta S,\delta q,\delta f,\delta h$ defined by:
\begin{eqnarray}
\delta S&\equiv&{2\over\gamma-1}{\delta c\over c}-{\delta\rho\over\rho},		\label{eq:def_dS} \\
\delta q&\equiv&\delta\left(\int ^r\frac{\cal L}{\rho v_r}dr'\right),   			  	\label{eq:def_dq}\\
\delta f&\equiv& v_r \delta v_r +\frac{2}{\gamma -1}c\delta c -\delta q,    		\label{eq:def_df}\\
\delta h&\equiv& \frac{\delta v_r}{v_r}+\frac{\delta \rho}{\rho},      			\label{eq:def_dh}
\end{eqnarray}
where $\delta$ denotes the Eulerian perturbation. As compared to the variables used by \citet{foglizzo07}, $\delta S$ and $\delta h$ are unchanged, the new variable $\delta q$ is related to $\delta K $ by $\delta K = l(l+1)\delta q$, and $\delta f$ is slightly changed to incorporate the perturbed heating $\delta q$. The usual physical quantities can be expressed as a function of these new variables as:
\begin{eqnarray}
{\delta v_r\over v}&=&{1\over 1-\M^2}\left(\delta h+\delta S-{\delta f\over c^2} - {\delta q \over c^2}\right),	
	\label{eq:dvr}\\
{\delta \rho\over\rho}&=&{1\over 1-\M^2}\left(-\M^2\delta h-\delta S+{\delta f + \delta q\over c^2}\right),
	\label{eq:drho}\\
{\delta P\over \gamma P}&=&{1\over 1-\M^2}\bigg\lbrack - \M^2\delta h-(1+(\gamma-1)\M^2){\delta S\over\gamma} \nonumber \\
		&&   + {\delta f + \delta q \over c^2}\bigg\rbrack,
	\label{eq:dp}\\
{\delta c^2\over c^2}&=&{\gamma-1\over 1-\M^2}\left({\delta f+ \delta q \over c^2}-\M^2{\delta h} - \M^2{\delta S}\right).
	\label{eq:dc2}
\end{eqnarray}
The velocity in non radial directions has a different spatial structure and is expressed as a function of angular derivative of the spherical harmonics in the following way:\begin{eqnarray}
\delta v_\theta &=& \frac{\delta f}{i\omega r} \frac{\p Y_l^m}{\p \theta}	,			\label{eq:vtheta}\\
\delta v_\phi &=& \frac{\delta f}{i\omega r \sin\theta} \frac{\p Y_l^m}{\p \phi}.		\label{eq:vtheta}
\end{eqnarray}

The advantage of this new formulation is that the differential system describing the perturbations becomes particularly compact (note that it is strictly equivalent to that of \citet{foglizzo07}):
\begin{eqnarray}
\frac{{\rm d} \delta f}{{\rm d} r}&=&\frac{i\omega c^2}{v_r(1-{\cal M}^2)}\bigg\{{\cal M}^2 \delta h - {\cal M}^2 \frac{\delta f}{c^2} \nonumber \\
		&&+[1+(\gamma -1){\cal M}^2 ]\frac{\delta S}{\gamma}-\frac{\delta q}{c^2}\bigg\},     	
	\label{eq:diff_df}\\
\frac{{\rm d} \delta h}{{\rm d} r}&=&\frac{i\omega}{v_r(1-{\cal M}^2)}\bigg\{\frac{\mu^2}{c^2}\delta f-{\cal M}^2 \delta h   \nonumber \\
		&&- \delta S +\frac{\delta q}{c^2}\bigg\},
	\label{eq:diff_dh}\\
\frac{{\rm d}\delta S}{{\rm d} r}&=&\frac{i\omega}{v_r}\delta S +\delta\left(\frac{\cal L}{Pv_r}\right),
	\label{eq:diff_dS}\\
\frac{{\rm d}\delta q}{{\rm d} r}&=&\frac{i\omega}{v_r}\delta q +\delta\left(\frac{\cal L}{\rho v_r}\right),
	\label{eq:diff_dq}
\end{eqnarray}
where $\mu$ is defined as in \citet{foglizzo07} by:
\begin{equation}
\mu^2 \equiv 1-\frac{l(l+1)c^2}{\omega^2r^2}(1-{\cal M}^2),
	\label{eq:def_mu}
\end{equation}
and the explicit expressions of $\delta({\cal L}/\rho v_r)$ and $\delta({\cal L}/P v_r)$ are given in \citet{foglizzo07} (Equations~(A7) and (A8)).

The boundary conditions at the shock are obtained by imposing the Rankine-Hugoniot relations for the perturbed quantities, which are written as:
\begin{eqnarray}
\delta f_{\rm sh}&=&i\omega v_{r,1} \Delta\zeta\left(1-\frac{v_{r,\rm sh}}{v_{r,1}}\right),
\label{eq:dfsh}\\
\delta h_{\rm sh}&=&-\frac{i\omega}{v_{r,\rm sh}}\Delta\zeta\left(1-\frac{v_{r,\rm sh}}{v_{r,1}}\right),
\label{eq:dhsh}\\
\frac{\delta S_{\rm sh}}{\gamma}&=&i\frac{\omega v_{r,1}}{c_{\rm sh}^2}\Delta\zeta\left(1-\frac{v_{r,\rm sh}}{v_{r,1}}\right)^2
-\frac{{\cal L}_{\rm sh}-{\cal L}_1}{\rho_{\rm sh} v_{r,\rm sh}}\frac{\Delta \zeta}{c_{\rm sh}^2}  \nonumber \\
		&& + \left(1-\frac{v_{r,\rm sh}}{v_{r,1}}\right)\frac{\Delta\zeta}{c_{\rm sh}^2}\left(\frac{2v_{r,1} v_{r,\rm sh}}{r_{\rm sh}}-\frac{{\rm d} \Phi}{{\rm d} r}\right),
\label{eq:dSsh}\\
\delta q_{\rm sh}&=&-\frac{{\cal L}_{\rm sh}-{\cal L}_1}{\rho_{\rm sh} v_{r,\rm sh}}\Delta\zeta,
\label{eq:dqsh}
\end{eqnarray}
where the subscripts '${\rm sh}$' and '1' refer to the values just below and above the shock, respectively; $\Delta \zeta$ is the radial displacement of the shock surface.

\bibliography{supernovae}

\bsp
\label{lastpage}

\end{document}